\documentclass[aps,english,nofootinbib,twocolumn]{revtex4-1}
\usepackage[T1]{fontenc}
\usepackage[utf8]{inputenc}
\usepackage{color}
\usepackage{babel}
\usepackage{pmboxdraw}
\usepackage{amstext}
\usepackage{graphicx}

\begin{document}

\title{Doped Graphene Quantum Dots UV-Vis Absorption Spectrum: A high-throughput
TDDFT study}
\author{\c{S}ener Özönder}
\affiliation{Department of Electrical-Electronics Engineering, Istinye University,
Istanbul, Turkey}
\email{Corresponding author: ozonder@umn.edu}
\author{Caner Ünlü}
\affiliation{Department of Chemistry, Istanbul Technical University, Istanbul,
Turkey}
\author{Cihat Güleryüz}
\affiliation{Department of Physics, Marmara University, Istanbul, Turkey}
\affiliation{Department of Opticianry, Alt\i nba\c{s} University, Istanbul, Turkey}
\author{Levent Trabzon}
\affiliation{Department of Mechanical Engineering, Istanbul Technical University,
Istanbul, Turkey}
\begin{abstract}
We report on time-dependent density functional theory (TDDFT) calculations
of the excited states of 63 different graphene quantum dots (GQDs)
in square shape with side lengths 1 nm, 1.5 nm and 2 nm. We investigate
the systematics and trends in the UV-Vis absorption spectra of these
GQDs, which are doped with elements B, N, O, S and P at dopant percentages
1.5\%, 3\%, 5\% and 7\%. The results show how the peaks in the UV
and visible parts of the spectrum as well as the total absorption
evolve in the chemical parameter space along the coordinates of size,
dopant type and dopant percentage. The absorption spectra calculated
here can be used to obtain particular GQD mixture proportions that
would yield a desired absorption profile such as flat absorption across
the whole visible spectrum or one that is locally peaked around a
chosen wavelength.
\end{abstract}
\maketitle

\section*{Introduction}

Graphene quantum dots (GQD) are two-dimensional, a few nanometer-sized
nanocrystals with tunable optical properties. Their applications are
ranging from solar cells to semiconductors as well as energy storage
and biomedical research \citep{review-Bak,review-henna,review-Xuewan}.
GQDs offer great functionality for light-harvesting and photoluminescence
applications since their optical properties can be tuned by changing
their size and chemically doping them with different elements. GQDs
also inherit all other useful properties of graphene such as low toxicity,
low cost and easy production and biocompatibility. 

Absorption spectrum of graphene determines its light-harvesting capacity
and it depends on the underlying electronic structure. Infinite graphene
crystal is a zero-band gap semi-metal, but when it is reduced to a
nanocrystal of a few nanometers, quantum confinement effects set in
and a band gap emerges. Also, chemical doping alters  graphene's
electronic structure and turns it into a p-type or n-type semiconductor
depending on the dopant type \citep{chemical-doping}. In the applications
of solar cells, quantum dot labeling, quantum dot enhanced photosynthesis
and optical sensors, absorption spectrum of the material needs to
be engineered in order it to be sensitive to the targeted part of
the spectrum.

Absorption spectra of selected GDQs with particular size, shape and
dopant type have been investigated in the past, however, a complete
systematic study considering the full chemical space of GDQs is currently
lacking. Here we report on a high-throughput scanning of GQDs of size
1-2 nm, of dopant elements B, N, O, S, P and of dopant percentages
from 0\% to 7 \% via time-dependent density functional theory (TDDFT)
calculations. The goal of this work is to chart the GQD chemical landscape
and to extract the physics on the systematics of how absorption spectrum
depends on the nanocrystal size and also how it is altered with chemical
doping. This information is necessary for a better spectrum engineering
of GQDs.

GQDs have been used in solar cells as an additive to better utilize
the UV part of the spectrum that is otherwise left unharvested in
conventional solar cells. Typically, GQDs possess $n-\pi$ and $n-\pi^{*}$
absorption bands which are located in the UV region of the light spectrum.
The absorption bands of GQDs can be controlled by manipulating either
the size or composition of GQDs through doping with heteroatoms. Graphene
is an excellent electron acceptor with mobility around $7\times10^{4}\text{ cm}^{2}\cdot\text{V}^{-1}\cdot\text{s}^{-1}$
and therefore has a great potential to improve efficiency of solar
cells as a charge carrier \citep{Luminscent}. GQDs, very small-sized
graphene fragments whose band-gap can be controlled, are already in
use for improving photovoltaic parameters of solar cells (photoconversion
efficiency, the peak power, the short-circuit current density, the
open circuit voltage and the fill factor) \citep{CanerDevelopment,Luminscent}.

\begin{figure*}[!t]
\includegraphics[width=1\textwidth]{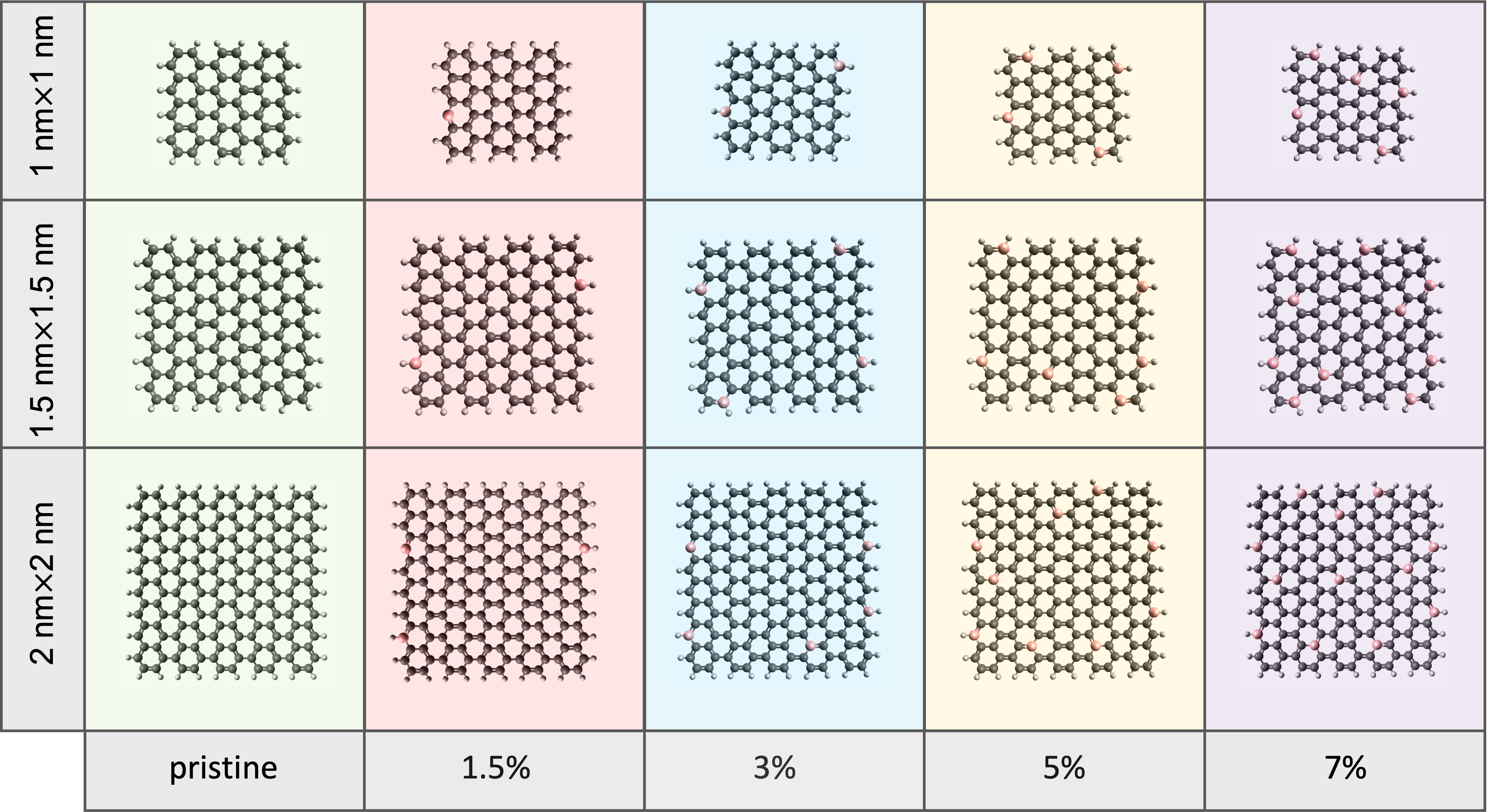}

\caption{Topology of the GQDs whose excited states and UV-Vis spectra are calculated. The carbon atoms are shown in gray whereas
the dopant elements are shown in pink. The circumferences are passivized
with hydrogen atoms. Pristine 1 nm, 1.5 nm and 2 nm GQDs (leftmost
column) are $\mathrm{C_{54}H_{22}}$, $\mathrm{C_{104}H_{32}}$ and
$\mathrm{C_{170}H_{46}}$, respectively.}

\label{GQDs-matrix}
\end{figure*}

In principle, GQDs can be produced through top-down or bottom-up synthesis
methods \citep{GQD,GQDschemApp}. However, controlling quality and
quantity of the dopant and the size of a GQD can be achieved through
bottom-up synthesis methods more precisely {\citep{GQDschemApp,BUDAK2021110577,GENCER2022108874}}.
GQDs can be synthesized through several different bottom-up synthesis
methods such as hydrothermal synthesis method, microwave assisted
synthesis method and solvothermal synthesis method \citep{GQDschemApp}.
Each bottom-up technique depends on incomplete carbonization of a
suitable carbon precursor and generally the carbon precursor is chosen
among biocompatible and easily affordable ones like citric acid, glucose,
etc. The size control of GQDs can be achieved
by controlling synthesis parameters such as temperature, pressure,
carbon precursors and solvent. Also, the composition
of GQDs can be controlled by addition of an extra heteroatom precursor
(N, B, S and P). As a result, the optical parameters
of GQDs can be manipulated via bottom-up synthesis techniques by controlling
the synthesis conditions and carbon or heteroatom precursors \citep{GQDschemApp}.

GQDs come in different shapes and dopant content and their  absorption
and emission properties are determined by these structural properties.
It is often not possible to estimate the optical properties of GQDs
from their molecular configurations by simple heuristic means, nor
is it possible to explore the optical properties of vast number of
GQD derivatives through laboratory synthesis. Also, prior knowledge
on the structure of a compound is needed to guide a chemist in what
to try in the lab. Density functional theory (DFT) provides such
guidance where vibrational, structural, electronic and optical properties
of a molecule or crystal can be calculated \emph{in silico}. In addition,
computational chemistry methods shed light into the underlying physical
mechanisms as well as possible effects that can be understood only
via simulations and are otherwise likely to be missed due to environmental
effects and errors in the measurement processes. Particularly for
exploring the optical properties such as absorptance and fluorescence
of molecules and crystals at a reasonable cost, time-dependent density
functional theory (TDDFT) has become the gold standard in recent years
\citep{benzene-JACQUEMIN,PARAC200311,tutorial-Jacquemin,Jacquemin-review,Accuracy-of-TdDFT}.
TDDFT can be used to calculate the excited states from which absorption
and emission spectra can be calculated, and it is used for discovering
and designing new compounds and as a complementary source of information
for verification and interpretation of the experimental results. 

There are several TDDFT studies in the literature that provide excited
states and absorption spectra of GQDs of particular size and dopant
content. Some of those past work focus on GDQs in specific shapes
such as triangle or hexagonal while some others focus on different
percentages of a single dopant element \citep{comparative-DFT,bursa-ytu,ABDELSALAM2018138,graphene-clusters,Absorption-Fluorescence,Ozfidan,pure-graphene-dft,nanoflakes,size-effect,C24-32-42-132,Fengdoped,oxidation}.
These individual studies do not adequately capture the mapping between
the various possible GQD structures and their absorption spectra.
In this work, we aim to fill this gap by calculating absorption spectra
of 63 different GQDs in the 3D parameter space (i) for side length
of 1, 1.5 and 2 nm, (ii) for dopants elements B, N, O, S and P, and
(iii) for dopant percentages 0\%, 1.5\%, 3\%, 5\% and 7\%. The results
can be used to find out the particular GDQ or mixture of them  which
will absorb the part of the spectrum the most as required by the specific
application in use. 

\begin{figure*}
\begin{centering}
\includegraphics[width=1\columnwidth]{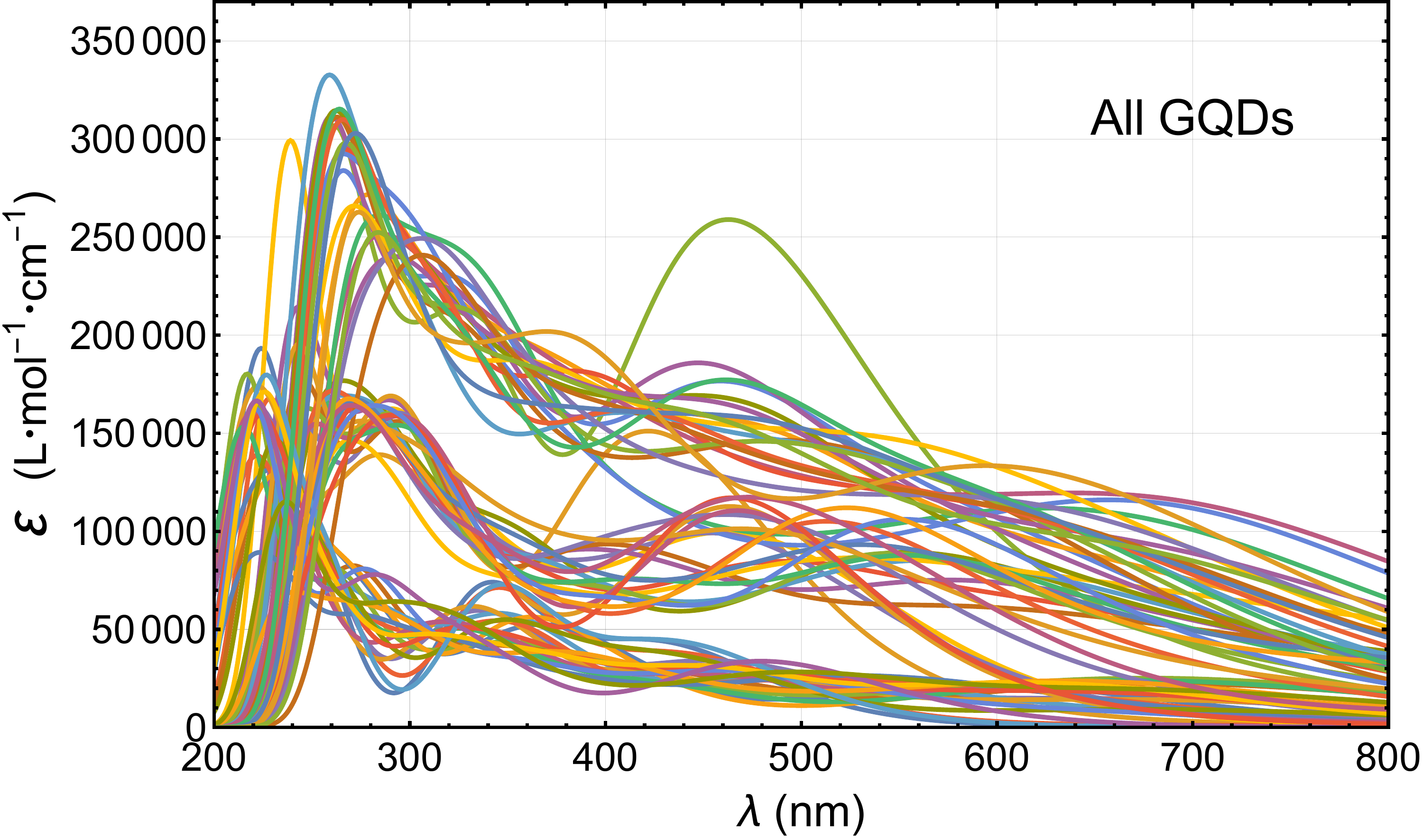}\includegraphics[width=1\columnwidth]{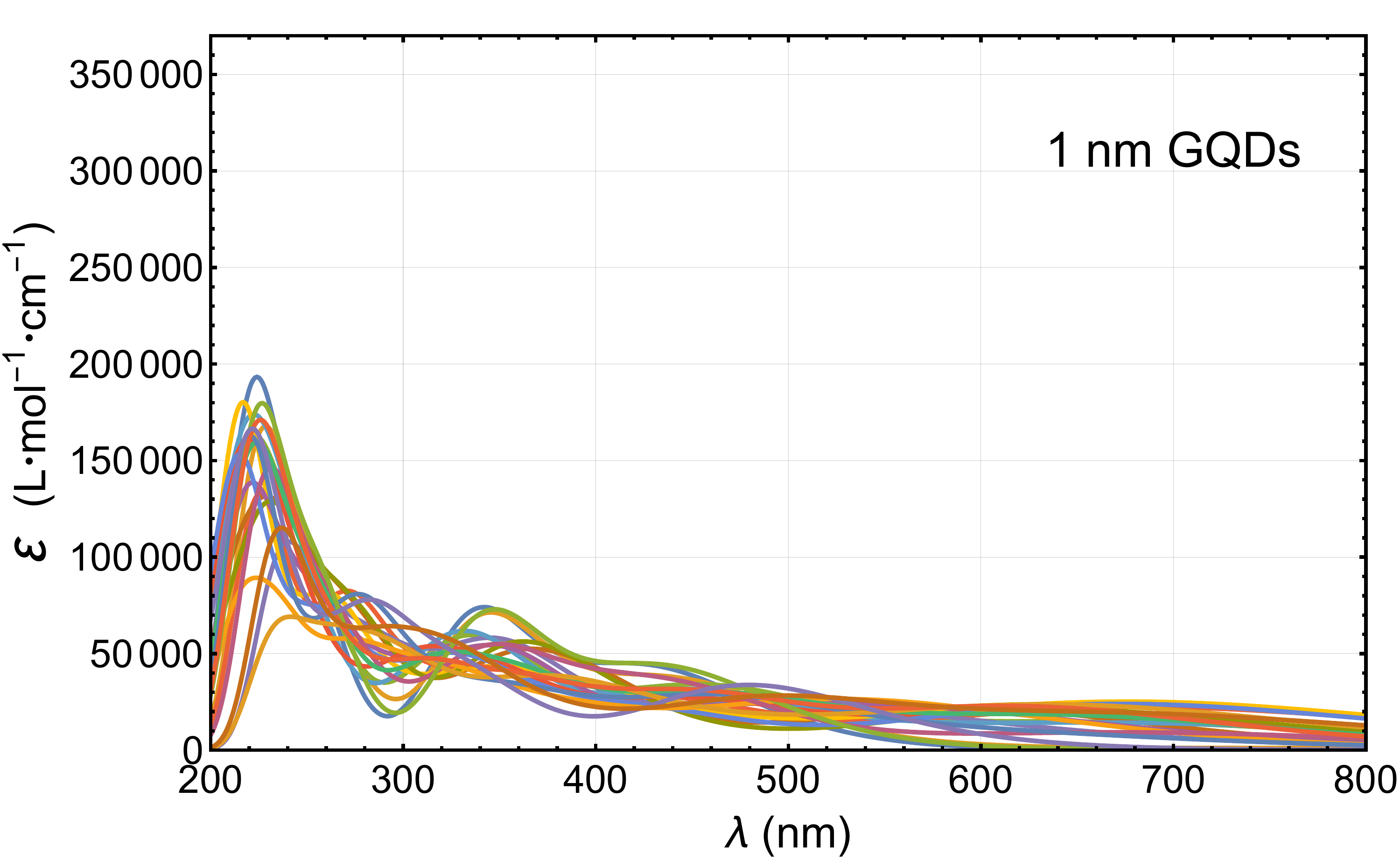}
\par\end{centering}
\begin{centering}
\includegraphics[width=1\columnwidth]{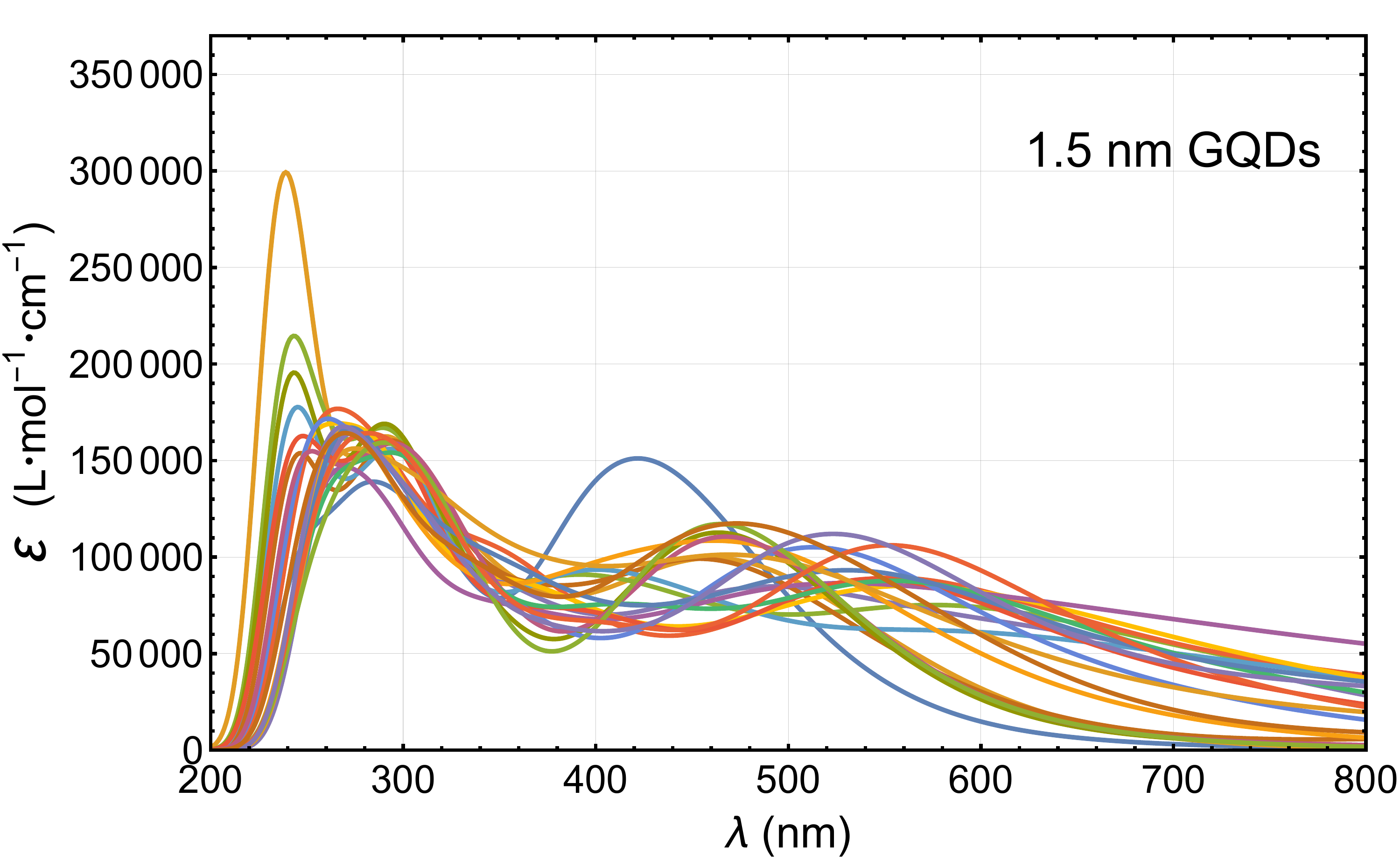}\includegraphics[width=1\columnwidth]{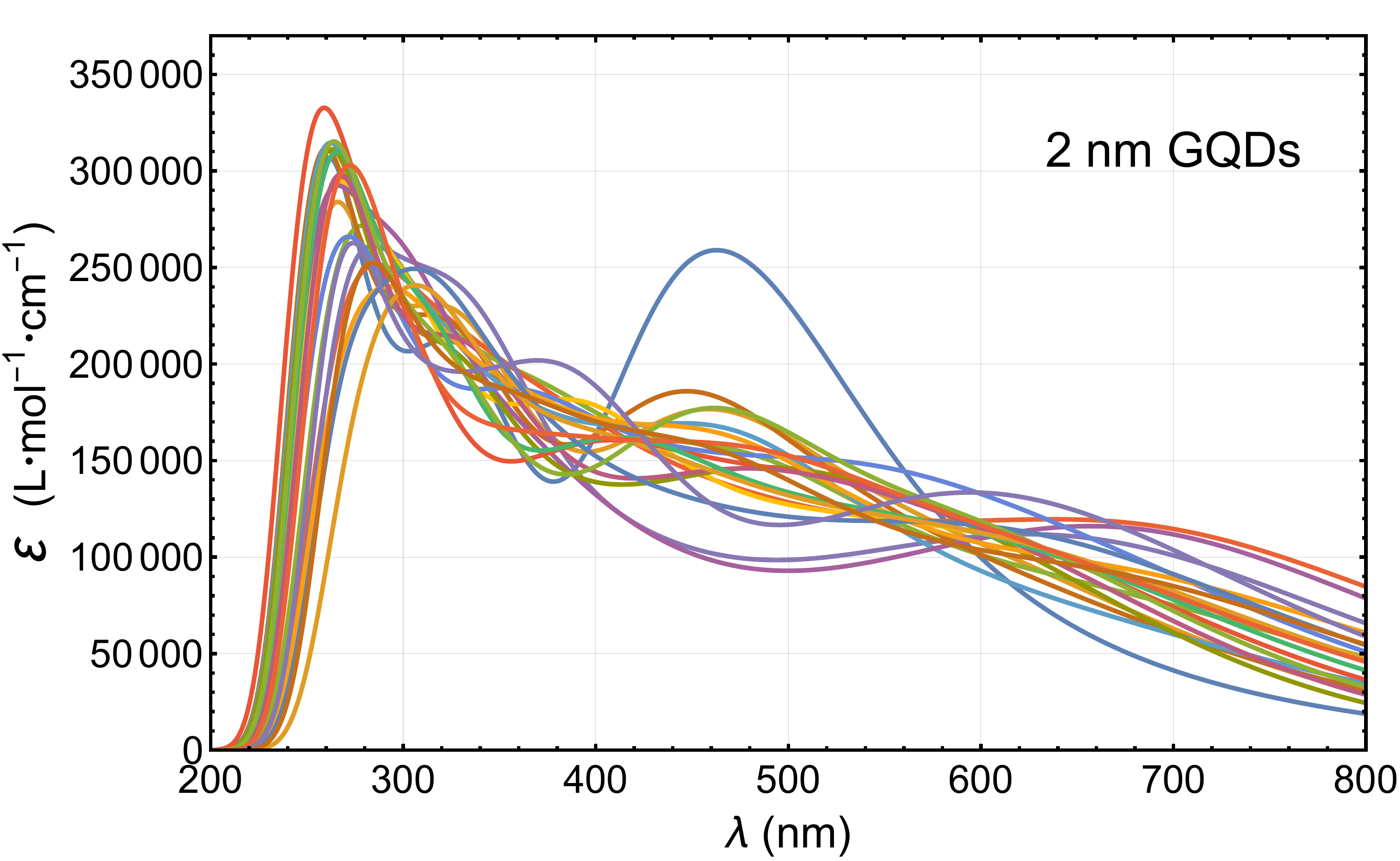}
\par\end{centering}
\caption{UV-Vis absorption spectra of GQDs. Top-left panel includes the spectra
of all 63 different GQDs, and the other panels have GQDs grouped with
respect to their sizes (side lengths). These spectra are calculated
with TDDFT with solvent being water. The size-dependent trends are
visible in the figures. The spectrum of each GQD with the labels of
size, dopant type and dopant percentage are provided separately in
Supporting Information.}

\label{selected-abs-spect}
\end{figure*}

\section*{Computational Details}

We calculate square-shaped graphene nanosheets with side lengths 1,
1.5 and 2 nm. The carbon atoms on the perimeter are passivized with
hydrogen atoms. In some cases, additional hydrogen atoms are added
depending on the dopant element type and percentage, in order to saturate
the free bonds and consequently ensure that the nanocrystal is charge
neutral and in $S=1$ singlet spin state. Fig. \ref{GQDs-matrix}
shows dopant locations for different dopant percentages for all three
sizes. 

Both geometry optimization and excited state calculations of GQDs
have been performed with Gaussian16 by using the hybrid functional
B3LYP with the basis set 6-31G(d). Past studies show that the model
B3LYP/6-31G(d) strikes the best balance between computational
cost and accuracy \citep{Absorption-Fluorescence,lithium,Jacquemin-review,SHOKUHIRAD2016176,RAD2015233} 

Water has been chosen as the solvent and it has been incorporated
in the calculations via polarizable continuum model (PCM) during both
geometry optimization and excited state calculations. During the optimization
evaluations, we also calculated the vibrational frequencies to ensure
the system is truly at the minimum of the potential energy surface.
In cases where the optimization ended up with negative frequencies,
we slightly distorted the atomic configuration in the direction of
those negative frequency vectors and rerun the optimization until
no negative frequencies remained. Once the optimized geometries were
obtained, vertical electronic excitation energies of each GQD were
calculated with the TDDFT method for the UV-visible part of the spectrum,
i.e., 1.6-5 eV (775-248 nm). For the hybrid functional B3LYP used,
the error in the excited state energies are expected to be in the
range of 0.20-0.25 eV \citep{tutorial-Jacquemin,Jacquemin-review}.

\begin{figure*}
\begin{centering}
\includegraphics[width=1\textwidth]{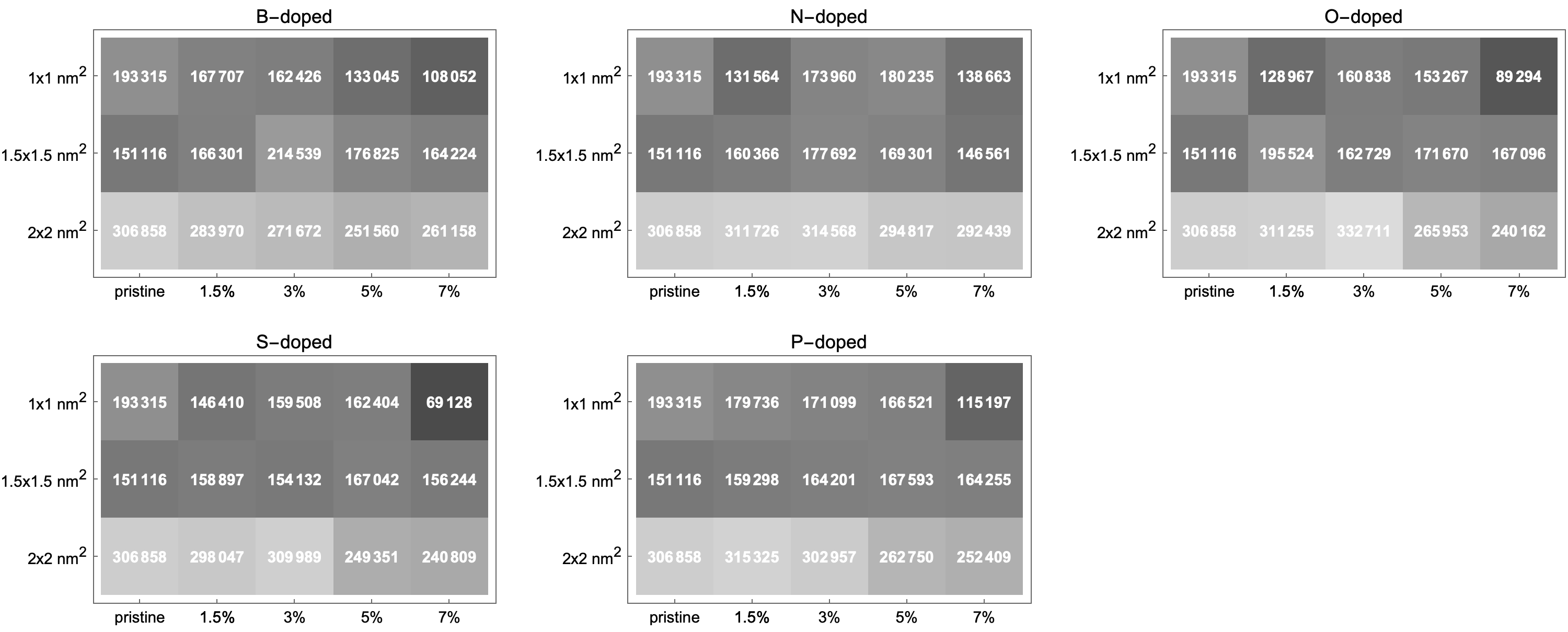}
\par\end{centering}
\caption{Peak values of extinction coefficient of pristine and doped GQDs calculated
via TDDFT. The labels at the top of each plot refer to the dopant
elements. For each dopant, a separate array plot with axes of size
and dopant percentage is given, and pristine GQDs are added to each
plot for comparison. The peak values of the extinction coefficients
given in each cell are usually positioned in the UV region with a
few exceptions, which can be examined in detail from the individual
spectra given in Supporting Information. The overall trend is that
the magnitude of the absorption (extinction coefficient) peak value
increases with increasing graphene size while it decreases with increasing
dopant percentage. }

\label{UV-max-values}
\end{figure*}

\section*{Results and Discussion}

The size, dopant type and dopant percentage change the electronic
structure of the GQDs, hence their absorption spectrum changes accordingly.
We convolved transition energies with gaussian distributions using
fwhm of $\sigma=$0.4 eV and used the oscillator strengths to calculate
the absorption spectrum for each GQD. Figure \ref{selected-abs-spect}
presents the absorption spectra of 63 GDQs plotted by using the TDDFT
excited state calculations in this work. The individual absorption
spectra of each GQDs are given in Supplementary Information (SI).
It is visible from the plots that the absorption spectrum, both the
magnitude and profile, depends on size, dopant type and dopant percentage.
The spectra turn out to be grouped in three different ``size bands'' with characteristic profiles.
Within a given size band, there is a trend that the
magnitude of the absorption decreases with increasing dopant percentage.
This is more apparent in the array plots given in Fig. \ref{UV-max-values}.
On the other hand, absorption (extinction coefficient) increases with
increasing GQD size. These trends look similar for all five dopant
elements used (B, N, O, S and P).

\begin{figure*}
\begin{centering}
\includegraphics[width=1\textwidth]{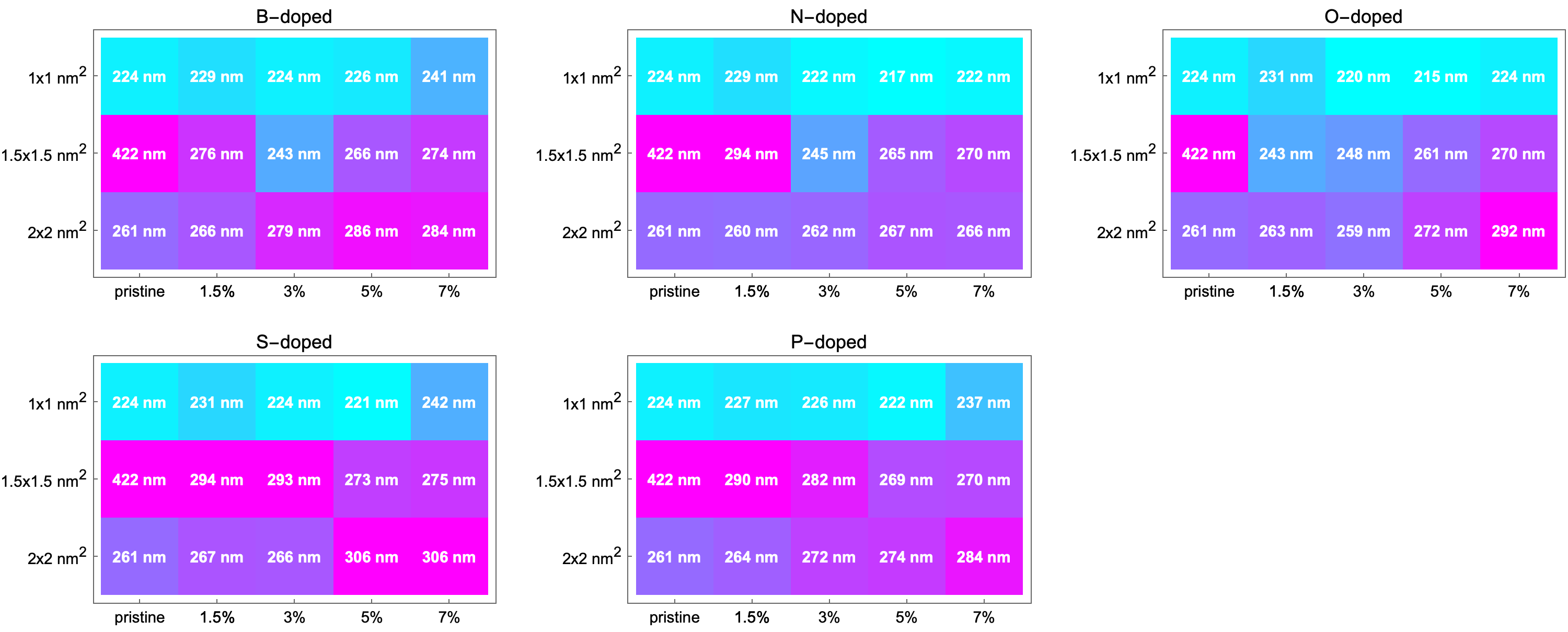}
\par\end{centering}
\caption{The positions of the absorption peaks in wavelength for each GQDs
in the UV-Vis region (200-800 nm) as calculated via TDDFT. The
letters refer to the dopant elements. Generally speaking, red-shifting
of the absorption peak is observed with increasing size as well as
increasing dopant percentage.}

\label{UV-max-wavelength}
\end{figure*}

Fig. \ref{UV-max-wavelength} shows the positions of the absorption
peaks in wavelength. In most cases, the peaks are red-shifted
with increasing graphene size. This effect
can be understood through the idea that larger size allows excitations
with longer wavelengths.

\begin{figure*}
\begin{centering}
\includegraphics[width=1\textwidth]{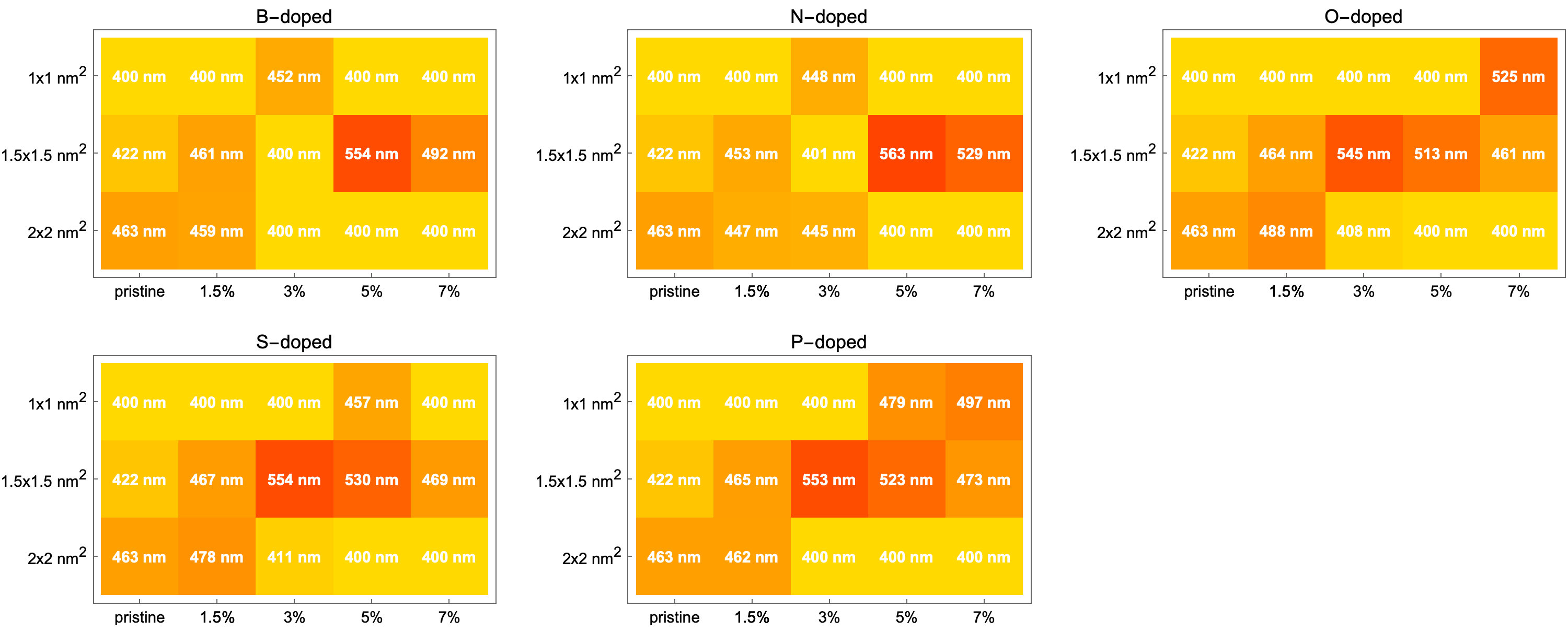}
\par\end{centering}
\caption{The positions of the absorption peaks in wavelength for each GQDs
in the visible region ($\lambda>400$ nm) as calculated via TDDFT.
The letters refer to the dopant elements. Relative red-shifting seems
to be occurring for the GQDs corresponding to the cells in the middle
of each array plot.}

\label{vis-max-wavelength}
\end{figure*}

The analysis of the peak positions can be narrowed down to only the
visible region if only the absorption of visible light is of interest
in a given application. Fig. \ref{vis-max-wavelength} shows the position
of the absorption peaks in wavelength in the visible region ($\lambda>400$
nm).

\begin{figure*}
\begin{centering}
\includegraphics[width=1\textwidth]{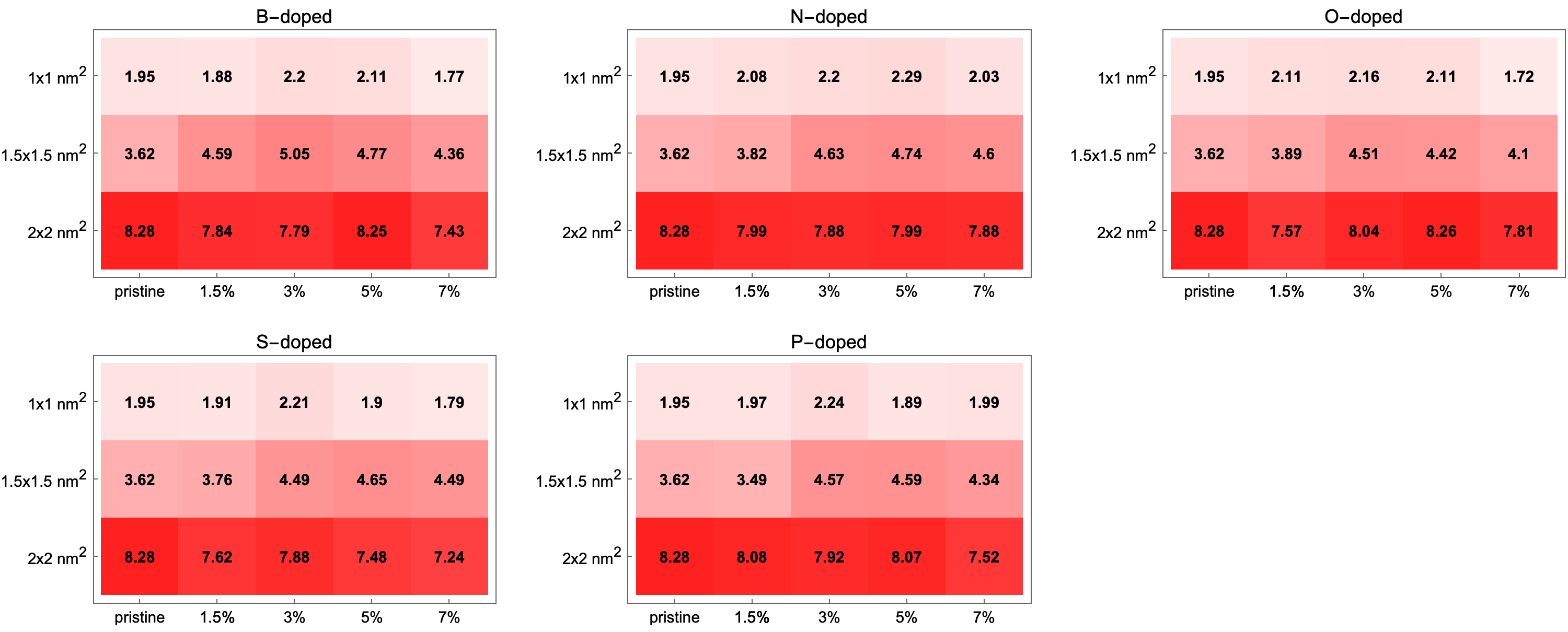}
\par\end{centering}
\caption{Total amount of absorption in the whole UV-Vis spectrum in units of
$\text{L}\cdot\text{mol}^{-1}$ found from the integral of the absorption
spectra that were calculated via TDDFT. Total absorption increases
with increasing GQD size. The letters refer to the dopant elements.
Total absorption grows with increasing GQD size. The change with dopant
type and dopant percentage exists but the latter does not have a simple
trend common to all cases.}

\label{areas}
\end{figure*}

So far, we've presented the systematics on how the absorption peak
values and their positions in wavelength change depending on size,
dopant type and dopant percentage. For certain applications, the matter
of interest may not be the absorption profiles but the total amount
of absorption in the whole UV-Vis spectrum. Figure \ref{areas} show
the integral of the absorption curves in units of $\text{L}\cdot\text{mol}^{-1}$.
The results show that total absorption over the whole UV-Vis spectrum
increases with increasing GQD size. It also changes with dopant type
and dopant percentage, but the direction of change varies depending
on the dopant and GQD size, so this should be examined case by case
from the array plots.

The results above demonstrates that the conjugated $\pi\text{-system}$
of the larger graphene nanosheets enable them to harvest light at
longer wavelengths which are missed by the smaller nanosheets most
likely due to quantum confinement effects in smaller ones. Thus, an
overall red shift in the absorption spectrum with increasing GQD size
is observed in the results. Similarly, larger graphene nanosheets
have higher total absorption since for them conjugations at both longer
and shorter wavelengths are available. The dopants disrupt this conjugated
$\pi\text{-system}$ and reduce the absorption at a given frequency
in comparison to that of the pristine graphene, however, the effect
of dopant percentage on the absorption profile is not drastic. Variation
in the dopant element type creates slight changes in the absorption
profile too, but the main determiner of the bands given in Fig. \ref{selected-abs-spect}
still remains to be the nanosheet size. Nevertheless, it shoud be
noted that the effect of dopant on optical parameters becomes more
significant for larger GQDs (above {1 nm $\times$ 1 nm}). So, the theoretical
results showed that doping GQDs with heteroatoms can be very useful
to manipulate optical parameters when size control is not possible
and GQDs with size around {1 nm $\times$ 1 nm} cannot be obtained.

\begin{figure}
\begin{centering}
\includegraphics[width=1\columnwidth]{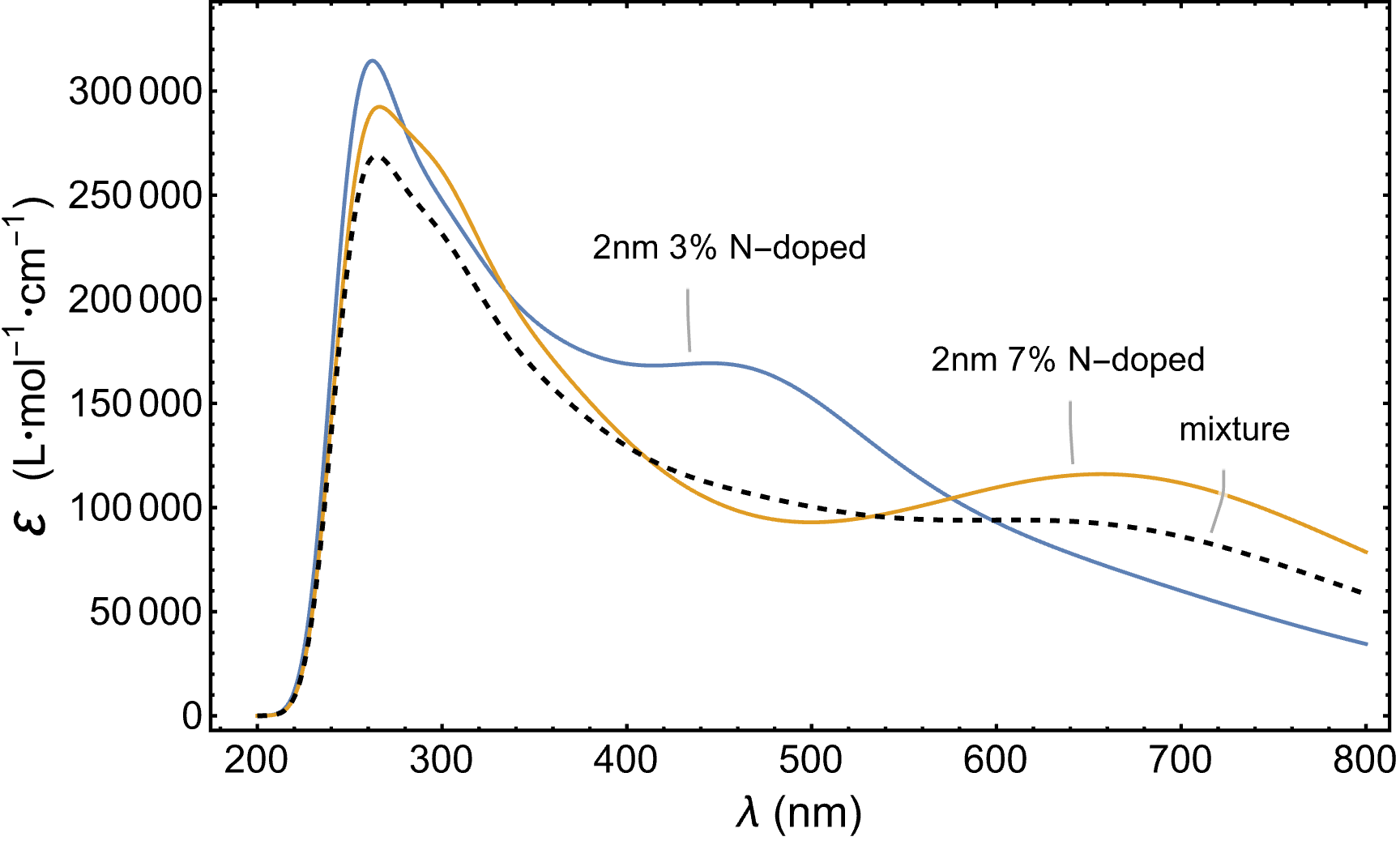}
\par\end{centering}
\caption{The absorption spectrum of the mixture, shown with a dashed line, of two
different GQDs producing a fairly flat absorption profile in the visible
spectrum (400-750 nm). The mixture here contains 28\% from ``2nm
3\% N-doped GQD'' and 62\% from ``2nm 7\% N-doped GQD''. }

\label{mixture}
\end{figure}

From the point of application, the outcome of this work can be
used for spectrum engineering for the situations where certain parts of
the UV-visible spectrum may be desired to have relatively more absorption.
For example, in the case of solar cell applications, one may be interested
in a specific mixture of the GQDs that has an absorption profile as
close as to a flat one across the whole visible spectrum. Figure \ref{mixture}
shows the spectrum of the mixture containing 28\% from ``2nm 3\%
N-doped GQD'' and 62\% from ``2nm 7\% N-doped GQD''. This specific
mixture gives rise to a fairly flat absorption spectrum in the visible
region.

As any computer simulation, TDDFT method has errors and these errors
are usually determined by the choice of the functional and basis set.
For example, for spatially extended Rydberg states, the functionals
wB97XD, CAM-B3LYP and M06-2X perform better for the complete profile
of the spectra whereas B3LYP may be enough as far as the peak positions
are concerned. \citep{Accuracy-of-TdDFT,PARAC200311,EOM-CCSD}. Furthermore,
to reduce the errors one may need to resort  to expensive  methods
such as EOM-CCSD \citep{EOM-CCSD}. However, a high-throughput investigation
of large nanocrystals with more than 200 atoms as in this work would
be prohibitively expensive, if not impossible, both in computational
resources as well as human workforce. While these considerations justify
the practical and necessary choice of TDDFT method at the B3LYP/6-31G(d)
level for the GQDs up to 2 nm here, the results presented should not
be seen as a precision study, instead, the general systematics and
trends in the results should be the main lesson to be taken here.

\section*{Conclusions and outlook}

We presented a TDDFT study of 63 different GQDs with systematically
varying size, dopant type and dopant percentage. The results suggested
visible trends in the peak properties as well as the general profile
in the absorption spectrum. The TDDFT calculations here shed light
on the systematics in the absorption properties of the nanometer-sized
graphene nanocrystals investigated in this work. A desired spectrum can be
obtained by mixing different GQDs with appropriate proportions, and
the spectra calculated in this work can be utilized to this end. This
study may be extended to codoping cases, different solvents, surface
functionals and other structural modifications of GQDs.

\section*{Acknowledgements}

\c{S}.Ö. thanks Kadir Diri for useful discussions. \c{S}.Ö. is supported
by TÜB\.{I}TAK under grant no. 120F354. This work is also supported
by Istanbul Technical University\textSFx Scientific Research Projects
Unit (ITU-BAP) {[}TOA-2019-42324{]}. Computing resources used in this
work were provided by the National Center for High Performance Computing
of Turkey (UHeM) under grant no. 1007872020.

\bibliographystyle{naturemag}
\bibliography{references}

\begin{thebibliography}{10}
\expandafter\ifx\csname url\endcsname\relax
  \def\url#1{\texttt{#1}}\fi
\expandafter\ifx\csname urlprefix\endcsname\relax\def\urlprefix{URL }\fi
\providecommand{\bibinfo}[2]{#2}
\providecommand{\eprint}[2][]{\url{#2}}

\bibitem{review-Bak}
\bibinfo{author}{Bak, S.}, \bibinfo{author}{Kim, D.} \& \bibinfo{author}{Lee,
  H.}
\newblock \bibinfo{title}{Graphene quantum dots and their possible energy
  applications: A review}.
\newblock \emph{\bibinfo{journal}{Current Applied Physics}}
  \textbf{\bibinfo{volume}{16}}, \bibinfo{pages}{1192--1201}
  (\bibinfo{year}{2016}).
\newblock \bibinfo{note}{Special Section on Nanostructure Physics and Materials
  Science at Center for Integrated Nanostructure Physics, Institute for Basic
  Science at Sungkyunkwan University}.

\bibitem{review-henna}
\bibinfo{author}{Henna, T.} \& \bibinfo{author}{Pramod, K.}
\newblock \bibinfo{title}{Graphene quantum dots redefine nanobiomedicine}.
\newblock \emph{\bibinfo{journal}{Materials Science and Engineering: C}}
  \textbf{\bibinfo{volume}{110}}, \bibinfo{pages}{110651}
  (\bibinfo{year}{2020}).

\bibitem{review-Xuewan}
\bibinfo{author}{Wang, X.} \emph{et~al.}
\newblock \bibinfo{title}{Heteroatom-doped graphene materials: syntheses{,}
  properties and applications}.
\newblock \emph{\bibinfo{journal}{Chem. Soc. Rev.}}
  \textbf{\bibinfo{volume}{43}}, \bibinfo{pages}{7067--7098}
  (\bibinfo{year}{2014}).

\bibitem{chemical-doping}
\bibinfo{author}{Liu, H.}, \bibinfo{author}{Liu, Y.} \& \bibinfo{author}{Zhu,
  D.}
\newblock \bibinfo{title}{Chemical doping of graphene}.
\newblock \emph{\bibinfo{journal}{J. Mater. Chem.}}
  \textbf{\bibinfo{volume}{21}}, \bibinfo{pages}{3335--3345}
  (\bibinfo{year}{2011}).

\bibitem{Luminscent}
\bibinfo{author}{Gupta, V.} \emph{et~al.}
\newblock \bibinfo{title}{Luminscent graphene quantum dots for organic
  photovoltaic devices}.
\newblock \emph{\bibinfo{journal}{Journal of the American Chemical Society}}
  \textbf{\bibinfo{volume}{133}}, \bibinfo{pages}{9960--9963}
  (\bibinfo{year}{2011}).
\newblock \bibinfo{note}{PMID: 21650464}.

\bibitem{CanerDevelopment}
\bibinfo{author}{Co\c{s}kun, Y.} \emph{et~al.}
\newblock \bibinfo{title}{Development of highly luminescent water-insoluble
  carbon dots by using calix[4]pyrrole as the carbon precursor and their
  potential application in organic solar cells}.
\newblock \emph{\bibinfo{journal}{ACS Omega}} \textbf{\bibinfo{volume}{0}},
  \bibinfo{pages}{null} (\bibinfo{year}{0}).

\bibitem{GQD}
\bibinfo{author}{Bacon, M.}, \bibinfo{author}{Bradley, S.~J.} \&
  \bibinfo{author}{Nann, T.}
\newblock \bibinfo{title}{Graphene quantum dots}.
\newblock \emph{\bibinfo{journal}{Particle \& Particle Systems
  Characterization}} \textbf{\bibinfo{volume}{31}}, \bibinfo{pages}{415--428}
  (\bibinfo{year}{2014}).

\bibitem{GQDschemApp}
\bibinfo{author}{Tian, P.}, \bibinfo{author}{Tang, L.}, \bibinfo{author}{Teng,
  K.} \& \bibinfo{author}{Lau, S.}
\newblock \bibinfo{title}{Graphene quantum dots from chemistry to
  applications}.
\newblock \emph{\bibinfo{journal}{Materials Today Chemistry}}
  \textbf{\bibinfo{volume}{10}}, \bibinfo{pages}{221--258}
  (\bibinfo{year}{2018}).

\bibitem{BUDAK2021110577}
\bibinfo{author}{Budak, E.} \& \bibinfo{author}{\"{U}nl\"{u}, C.}
\newblock \bibinfo{title}{Boron regulated dual emission in b, n doped graphene
  quantum dots}.
\newblock \emph{\bibinfo{journal}{Optical Materials}}
  \textbf{\bibinfo{volume}{111}}, \bibinfo{pages}{110577}
  (\bibinfo{year}{2021}).

\bibitem{GENCER2022108874}
\bibinfo{author}{Gencer, O.}, \bibinfo{author}{\"{O}mer Faruk~\c{C}even} \&
  \bibinfo{author}{\"{U}nl\"{u}, C.}
\newblock \bibinfo{title}{Triggering excitation independent fluorescence in
  zinc(ii) incorporated carbon dots: Surface passivation of carbon dots with
  zinc(ii) ions by microwave assisted synthesis methods}.
\newblock \emph{\bibinfo{journal}{Diamond and Related Materials}}
  \textbf{\bibinfo{volume}{123}}, \bibinfo{pages}{108874}
  (\bibinfo{year}{2022}).

\bibitem{benzene-JACQUEMIN}
\bibinfo{author}{Jacquemin, D.} \emph{et~al.}
\newblock \bibinfo{title}{Absorption and emission spectra in gas-phase and
  solution using td-dft: Formaldehyde and benzene as case studies}.
\newblock \emph{\bibinfo{journal}{Chemical Physics Letters}}
  \textbf{\bibinfo{volume}{421}}, \bibinfo{pages}{272--276}
  (\bibinfo{year}{2006}).

\bibitem{PARAC200311}
\bibinfo{author}{Parac, M.} \& \bibinfo{author}{Grimme, S.}
\newblock \bibinfo{title}{A tddft study of the lowest excitation energies of
  polycyclic aromatic hydrocarbons}.
\newblock \emph{\bibinfo{journal}{Chemical Physics}}
  \textbf{\bibinfo{volume}{292}}, \bibinfo{pages}{11--21}
  (\bibinfo{year}{2003}).

\bibitem{tutorial-Jacquemin}
\bibinfo{author}{Adamo, C.} \& \bibinfo{author}{Jacquemin, D.}
\newblock \bibinfo{title}{The calculations of excited-state properties with
  time-dependent density functional theory}.
\newblock \emph{\bibinfo{journal}{Chem. Soc. Rev.}}
  \textbf{\bibinfo{volume}{42}}, \bibinfo{pages}{845--856}
  (\bibinfo{year}{2013}).

\bibitem{Jacquemin-review}
\bibinfo{author}{Escudero, D.}, \bibinfo{author}{Laurent, A.~D.} \&
  \bibinfo{author}{Jacquemin, D.}
\newblock \emph{\bibinfo{title}{Time-Dependent Density Functional Theory: A
  Tool to Explore Excited States}}, \bibinfo{pages}{927--961}
  (\bibinfo{publisher}{Springer International Publishing},
  \bibinfo{address}{Cham}, \bibinfo{year}{2017}).

\bibitem{Accuracy-of-TdDFT}
\bibinfo{author}{Miyahara, T.} \& \bibinfo{author}{Nakatsuji, H.}
\newblock \bibinfo{title}{Accuracy of td-dft in the ultraviolet and circular
  dichroism spectra of deoxyguanosine and uridine}.
\newblock \emph{\bibinfo{journal}{The Journal of Physical Chemistry A}}
  \textbf{\bibinfo{volume}{122}}, \bibinfo{pages}{100--118}
  (\bibinfo{year}{2018}).
\newblock \bibinfo{note}{PMID: 29190101}.

\bibitem{comparative-DFT}
\bibinfo{author}{Jyoti~Tyagi, R.~K., Lekha~Sharma}.
\newblock \bibinfo{title}{Graphene and doped graphene: A comparative dft
  study}.
\newblock \emph{\bibinfo{journal}{Advanced Materials Letters}}
  \textbf{\bibinfo{volume}{10}}, \bibinfo{pages}{484--490}
  (\bibinfo{year}{2019}).

\bibitem{bursa-ytu}
\bibinfo{author}{Kayk{\i}larl{\i}, C.}, \bibinfo{author}{Uzunsoy, D.},
  \bibinfo{author}{Parmak, E. D.~{\c{S}}.}, \bibinfo{author}{Fellah, M.~F.} \&
  \bibinfo{author}{\"{O}zgen {\c{C}}olak~{\c{C}}ak{\i}r}.
\newblock \bibinfo{title}{Boron and nitrogen doping in graphene: an
  experimental and density functional theory ({DFT}) study}.
\newblock \emph{\bibinfo{journal}{Nano Express}} \textbf{\bibinfo{volume}{1}},
  \bibinfo{pages}{010027} (\bibinfo{year}{2020}).

\bibitem{ABDELSALAM2018138}
\bibinfo{author}{Abdelsalam, H.}, \bibinfo{author}{Elhaes, H.} \&
  \bibinfo{author}{Ibrahim, M.~A.}
\newblock \bibinfo{title}{Tuning electronic properties in graphene quantum dots
  by chemical functionalization: Density functional theory calculations}.
\newblock \emph{\bibinfo{journal}{Chemical Physics Letters}}
  \textbf{\bibinfo{volume}{695}}, \bibinfo{pages}{138--148}
  (\bibinfo{year}{2018}).

\bibitem{graphene-clusters}
\bibinfo{author}{Saha, B.} \& \bibinfo{author}{Bhattacharyya, P.~K.}
\newblock \bibinfo{title}{Understanding reactivity{,} aromaticity and
  absorption spectra of carbon cluster mimic to graphene: a dft study}.
\newblock \emph{\bibinfo{journal}{RSC Adv.}} \textbf{\bibinfo{volume}{6}},
  \bibinfo{pages}{79768--79780} (\bibinfo{year}{2016}).

\bibitem{Absorption-Fluorescence}
\bibinfo{author}{Zhao, M.}, \bibinfo{author}{Yang, F.}, \bibinfo{author}{Xue,
  Y.}, \bibinfo{author}{Xiao, D.} \& \bibinfo{author}{Guo, Y.}
\newblock \bibinfo{title}{A time-dependent dft study of the absorption and
  fluorescence properties of graphene quantum dots}.
\newblock \emph{\bibinfo{journal}{ChemPhysChem}} \textbf{\bibinfo{volume}{15}},
  \bibinfo{pages}{950--957} (\bibinfo{year}{2014}).

\bibitem{Ozfidan}
\bibinfo{author}{\"{O}zfidan, I.}, \bibinfo{author}{Güçlü, A.~D.},
  \bibinfo{author}{Korkusinski, M.} \& \bibinfo{author}{Hawrylak, P.}
\newblock \bibinfo{title}{Theory of optical properties of graphene quantum
  dots}.
\newblock \emph{\bibinfo{journal}{physica status solidi (RRL) - Rapid Research
  Letters}} \textbf{\bibinfo{volume}{10}}, \bibinfo{pages}{102--110}
  (\bibinfo{year}{2016}).

\bibitem{pure-graphene-dft}
\bibinfo{author}{Chopra, S.} \& \bibinfo{author}{Maidich, L.}
\newblock \bibinfo{title}{Optical properties of pure graphene in various forms:
  a time dependent density functional theory study}.
\newblock \emph{\bibinfo{journal}{RSC Adv.}} \textbf{\bibinfo{volume}{4}},
  \bibinfo{pages}{50606--50613} (\bibinfo{year}{2014}).

\bibitem{nanoflakes}
\bibinfo{author}{Lin, C.-K.}
\newblock \bibinfo{title}{Theoretical study of nitrogen-doped graphene
  nanoflakes: Stability and spectroscopy depending on dopant types and flake
  sizes}.
\newblock \emph{\bibinfo{journal}{Journal of Computational Chemistry}}
  \textbf{\bibinfo{volume}{39}}, \bibinfo{pages}{1387--1397}
  (\bibinfo{year}{2018}).

\bibitem{size-effect}
\bibinfo{author}{Zhang, P.} \emph{et~al.}
\newblock \bibinfo{title}{Size effect of oxygen reduction reaction on
  nitrogen-doped graphene quantum dots}.
\newblock \emph{\bibinfo{journal}{RSC Adv.}} \textbf{\bibinfo{volume}{8}},
  \bibinfo{pages}{531--536} (\bibinfo{year}{2018}).

\bibitem{C24-32-42-132}
\bibinfo{author}{Feng, J.}, \bibinfo{author}{Dong, H.}, \bibinfo{author}{Yu,
  L.} \& \bibinfo{author}{Dong, L.}
\newblock \bibinfo{title}{The optical and electronic properties of graphene
  quantum dots with oxygen-containing groups: a density functional theory
  study}.
\newblock \emph{\bibinfo{journal}{J. Mater. Chem. C}}
  \textbf{\bibinfo{volume}{5}}, \bibinfo{pages}{5984--5993}
  (\bibinfo{year}{2017}).

\bibitem{Fengdoped}
\bibinfo{author}{Feng, J.} \emph{et~al.}
\newblock \bibinfo{title}{Theoretical study on the optical and electronic
  properties of graphene quantum dots doped with heteroatoms}.
\newblock \emph{\bibinfo{journal}{Phys. Chem. Chem. Phys.}}
  \textbf{\bibinfo{volume}{20}}, \bibinfo{pages}{15244--15252}
  (\bibinfo{year}{2018}).

\bibitem{oxidation}
\bibinfo{author}{Chen, S.}, \bibinfo{author}{Ullah, N.}, \bibinfo{author}{Wang,
  T.} \& \bibinfo{author}{Zhang, R.}
\newblock \bibinfo{title}{Tuning the optical properties of graphene quantum
  dots by selective oxidation: a theoretical perspective}.
\newblock \emph{\bibinfo{journal}{J. Mater. Chem. C}}
  \textbf{\bibinfo{volume}{6}}, \bibinfo{pages}{6875--6883}
  (\bibinfo{year}{2018}).

\bibitem{lithium}
\bibinfo{author}{Ajeel, F.~N.}, \bibinfo{author}{Mohammed, M.~H.} \&
  \bibinfo{author}{Khudhair, A.~M.}
\newblock \bibinfo{title}{Effects of lithium impurities on electronic and
  optical properties of graphene nanoflakes: A dft-tddft study}.
\newblock \emph{\bibinfo{journal}{Chinese Journal of Physics}}
  \textbf{\bibinfo{volume}{58}}, \bibinfo{pages}{109--116}
  (\bibinfo{year}{2019}).

\bibitem{SHOKUHIRAD2016176}
\bibinfo{author}{{Shokuhi Rad}, A.}, \bibinfo{author}{Esfahanian, M.},
  \bibinfo{author}{Maleki, S.} \& \bibinfo{author}{Gharati, G.}
\newblock \bibinfo{title}{Application of carbon nanostructures toward so2 and
  so3 adsorption: a comparison between pristine graphene and n-doped graphene
  by dft calculations}.
\newblock \emph{\bibinfo{journal}{Journal of Sulfur Chemistry}}
  \textbf{\bibinfo{volume}{37}}, \bibinfo{pages}{176--188}
  (\bibinfo{year}{2016}).

\bibitem{RAD2015233}
\bibinfo{author}{Rad, A.~S.} \& \bibinfo{author}{Kashani, O.~R.}
\newblock \bibinfo{title}{Adsorption of acetyl halide molecules on the surface
  of pristine and al-doped graphene: Ab initio study}.
\newblock \emph{\bibinfo{journal}{Applied Surface Science}}
  \textbf{\bibinfo{volume}{355}}, \bibinfo{pages}{233--241}
  (\bibinfo{year}{2015}).

\bibitem{EOM-CCSD}
\bibinfo{author}{Acharya, A.}, \bibinfo{author}{Chaudhuri, S.} \&
  \bibinfo{author}{Batista, V.~S.}
\newblock \bibinfo{title}{Can tddft describe excited electronic states of
  naphthol photoacids? a closer look with eom-ccsd}.
\newblock \emph{\bibinfo{journal}{Journal of Chemical Theory and Computation}}
  \textbf{\bibinfo{volume}{14}}, \bibinfo{pages}{867--876}
  (\bibinfo{year}{2018}).

\end{thebibliography}

%\pagebreak

\vspace{3cm}
\textbf{\Large{}Supplementary Information for Doped Graphene Quantum
Dots UV-Vis Absorption Spectrum: A high-throughput TDDFT study}

%\section*{Supplementary Information}

\begin{figure*}
\hspace*{-1cm}
\includegraphics[width=18cm,keepaspectratio]{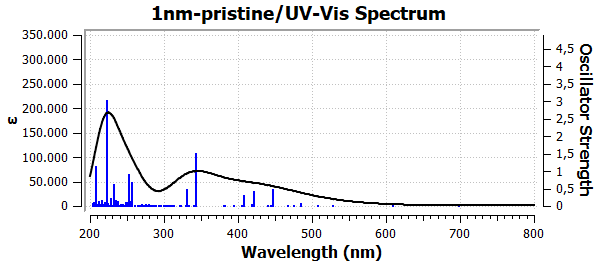} 
\end{figure*}

\begin{figure*}
\hspace*{-1cm}
\includegraphics[width=18cm,keepaspectratio]{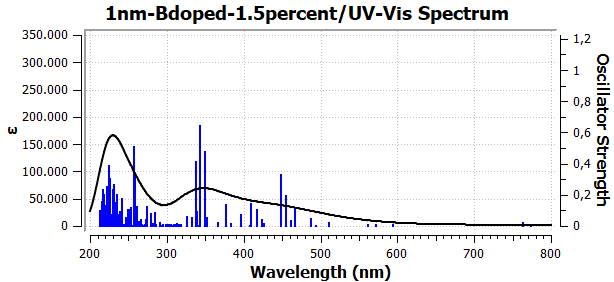}  
\end{figure*}

\begin{figure*}
\hspace*{-1cm}
\includegraphics[width=18cm,keepaspectratio]{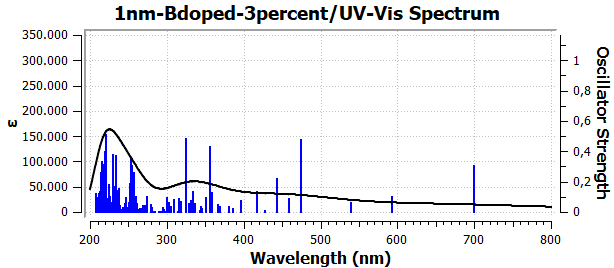} 
\end{figure*}

\begin{figure*}
\hspace*{-1cm}
\includegraphics[width=18cm,keepaspectratio]{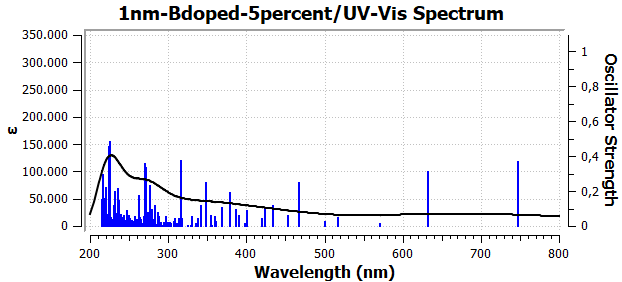} 
\end{figure*}

\begin{figure*}
\hspace*{-1cm}
\includegraphics[width=18cm,keepaspectratio]{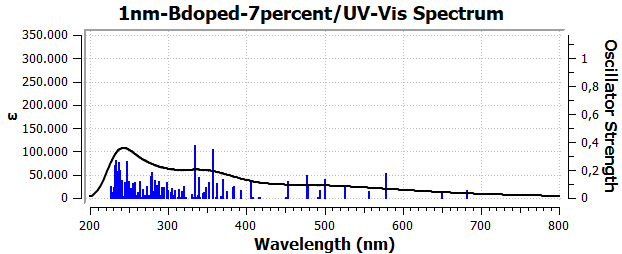} 
\end{figure*}

\begin{figure*}
\hspace*{-1cm}
\includegraphics[width=18cm,keepaspectratio]{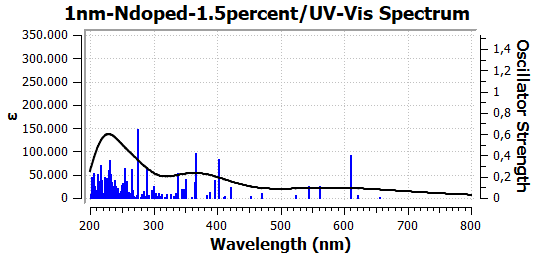} 
\end{figure*}

\begin{figure*}
\hspace*{-1cm}
\includegraphics[width=18cm,keepaspectratio]{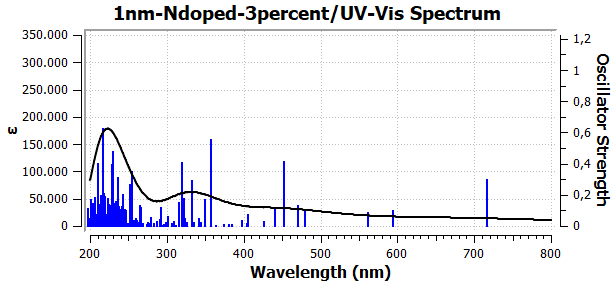} 
\end{figure*}

\begin{figure*}
\hspace*{-1cm}
\includegraphics[width=18cm,keepaspectratio]{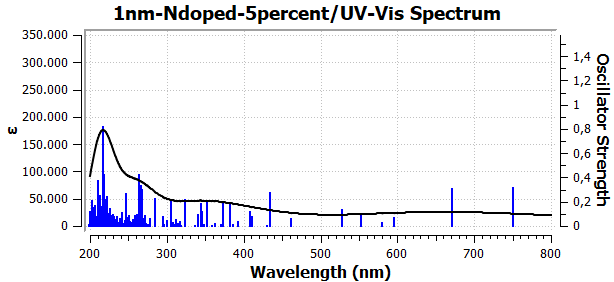} 
\end{figure*}

\begin{figure*}
\hspace*{-1cm}
\includegraphics[width=18cm,keepaspectratio]{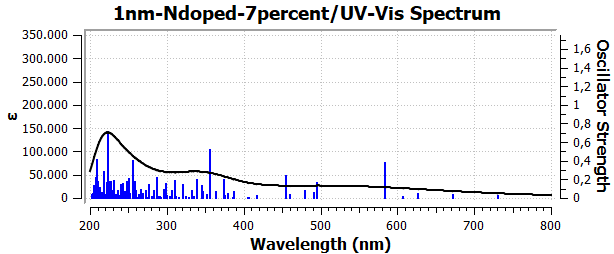} 
\end{figure*}

\begin{figure*}
\hspace*{-1cm}
\includegraphics[width=18cm,keepaspectratio]{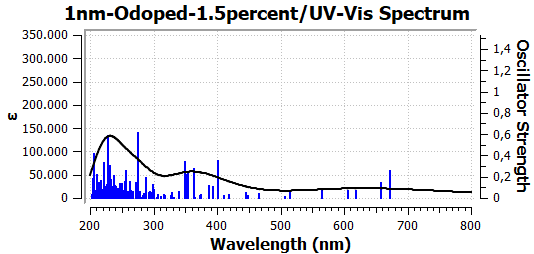} 
\end{figure*}

\begin{figure*}
\hspace*{-1cm}
\includegraphics[width=18cm,keepaspectratio]{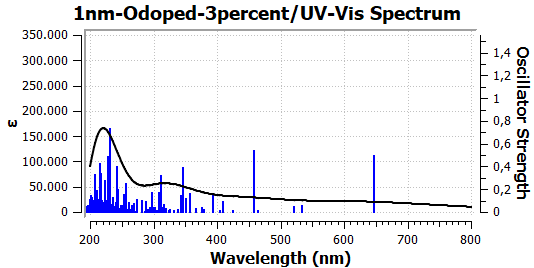} 
\end{figure*}

\begin{figure*}
\hspace*{-1cm}
\includegraphics[width=18cm,keepaspectratio]{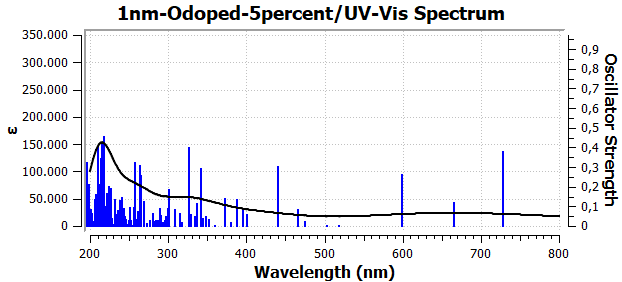} 
\end{figure*}

\begin{figure*}
\hspace*{-1cm}
\includegraphics[width=18cm,keepaspectratio]{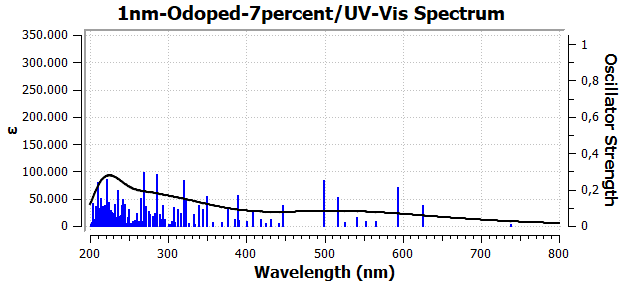} 
\end{figure*}

\begin{figure*}
\hspace*{-1cm}
\includegraphics[width=18cm,keepaspectratio]{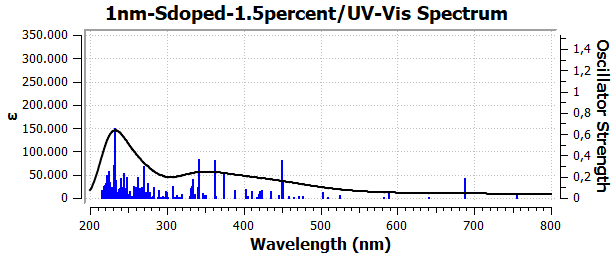} 
\end{figure*}

\begin{figure*}
\hspace*{-1cm}
\includegraphics[width=18cm,keepaspectratio]{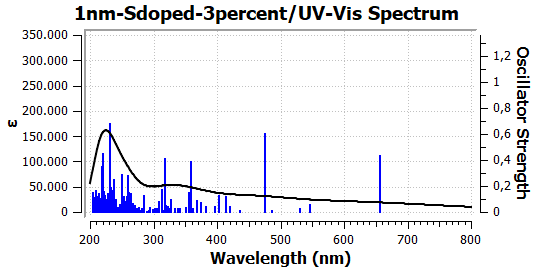} 
\end{figure*}

\begin{figure*}
\hspace*{-1cm}
\includegraphics[width=18cm,keepaspectratio]{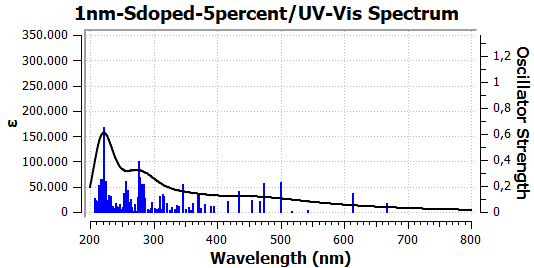} 
\end{figure*}

\begin{figure*}
\hspace*{-1cm}
\includegraphics[width=18cm,keepaspectratio]{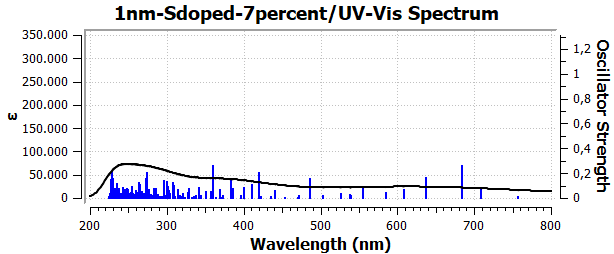} 
\end{figure*}

\begin{figure*}
\hspace*{-1cm}
\includegraphics[width=18cm,keepaspectratio]{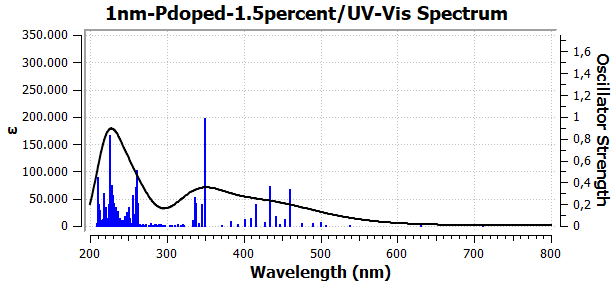} 
\end{figure*}

\begin{figure*}
\hspace*{-1cm}
\includegraphics[width=18cm,keepaspectratio]{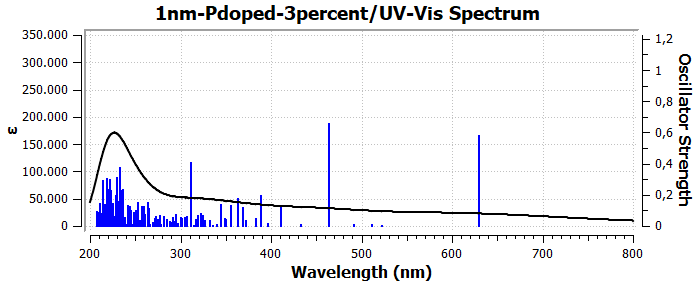} 
\end{figure*}

\begin{figure*}
\hspace*{-1cm}
\includegraphics[width=18cm,keepaspectratio]{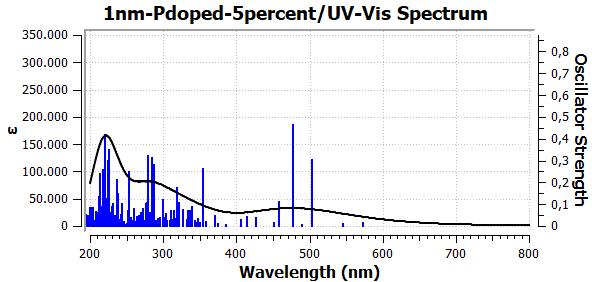} 
\end{figure*}

\begin{figure*}
\hspace*{-1cm}
\includegraphics[width=18cm,keepaspectratio]{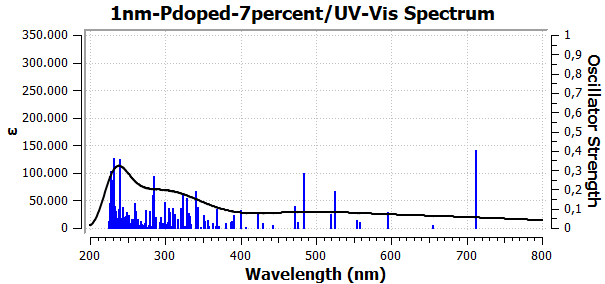} 
\end{figure*}

%1.5nm

\begin{figure*}
\hspace*{-1cm}
\includegraphics[width=18cm,keepaspectratio]{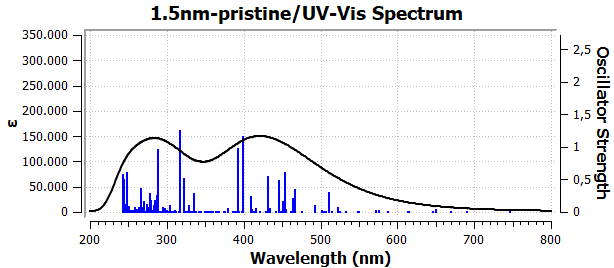} 
\end{figure*}

\begin{figure*}
\hspace*{-1cm}
\includegraphics[width=18cm,keepaspectratio]{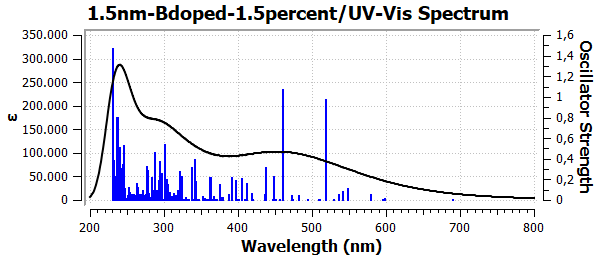} 
\end{figure*}

\begin{figure*}
\hspace*{-1cm}
\includegraphics[width=18cm,keepaspectratio]{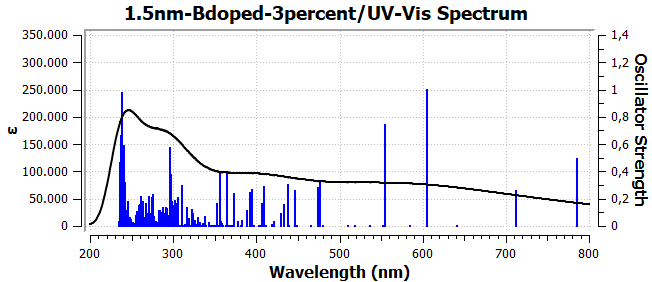} 
\end{figure*}

\begin{figure*}
\hspace*{-1cm}
\includegraphics[width=18cm,keepaspectratio]{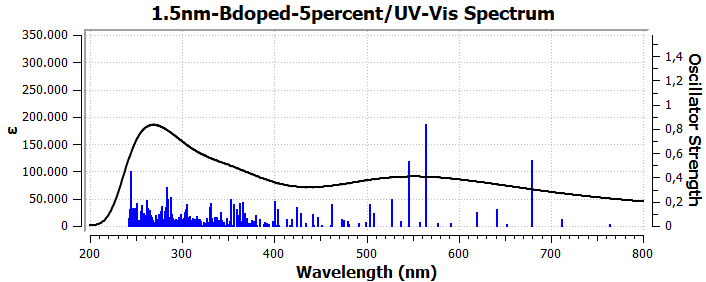} 
\end{figure*}

\begin{figure*}
\hspace*{-1cm}
\includegraphics[width=18cm,keepaspectratio]{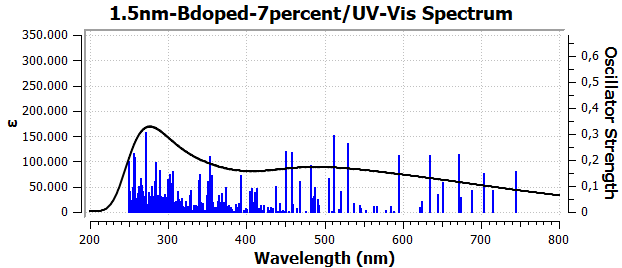} 
\end{figure*}

\begin{figure*}
\hspace*{-1cm}
\includegraphics[width=18cm,keepaspectratio]{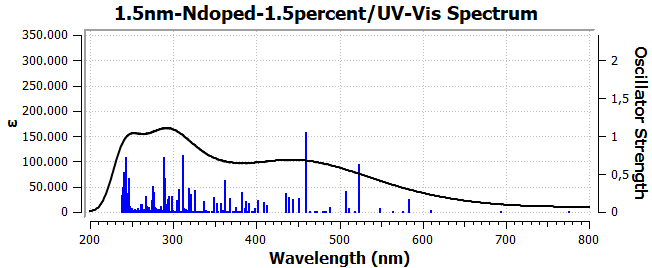} 
\end{figure*}

\begin{figure*}
\hspace*{-1cm}
\includegraphics[width=18cm,keepaspectratio]{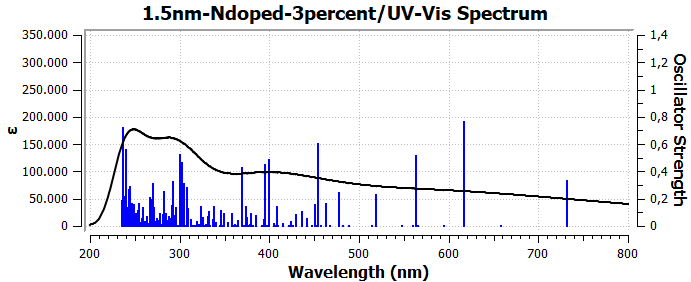} 
\end{figure*}

\begin{figure*}
\hspace*{-1cm}
\includegraphics[width=18cm,keepaspectratio]{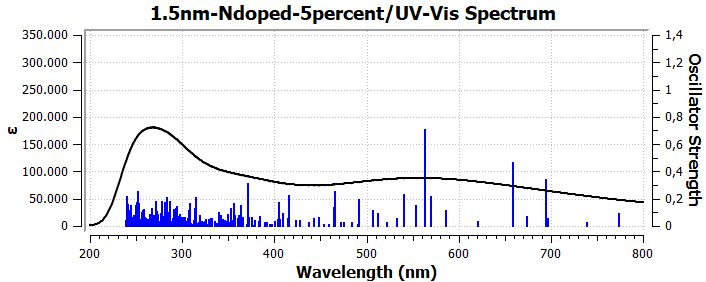} 
\end{figure*}

\begin{figure*}
\hspace*{-1cm}
\includegraphics[width=18cm,keepaspectratio]{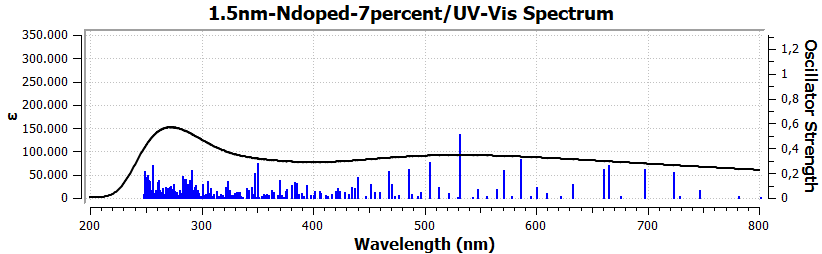} 
\end{figure*}

\begin{figure*}
\hspace*{-1cm}
\includegraphics[width=18cm,keepaspectratio]{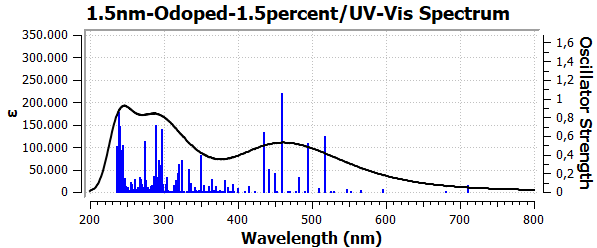} 
\end{figure*}

\begin{figure*}
\hspace*{-1cm}
\includegraphics[width=18cm,keepaspectratio]{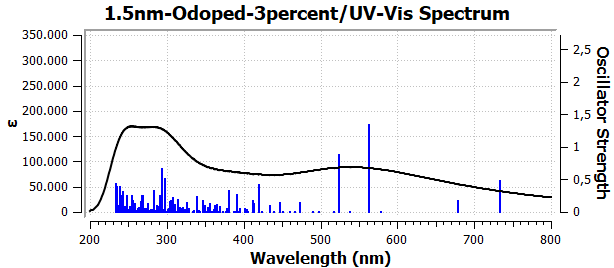} 
\end{figure*}

\begin{figure*}
\hspace*{-1cm}
\includegraphics[width=18cm,keepaspectratio]{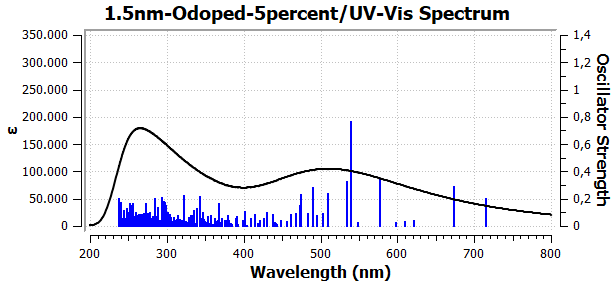} 
\end{figure*}

\begin{figure*}
\hspace*{-1cm}
\includegraphics[width=18cm,keepaspectratio]{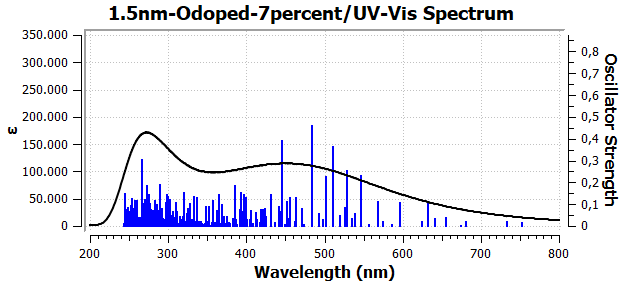} 
\end{figure*}

\begin{figure*}
\hspace*{-1cm}
\includegraphics[width=18cm,keepaspectratio]{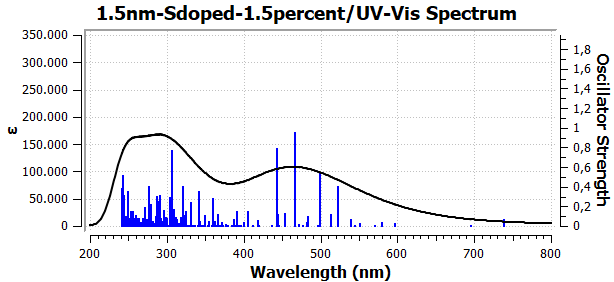} 
\end{figure*}

\begin{figure*}
\hspace*{-1cm}
\includegraphics[width=18cm,keepaspectratio]{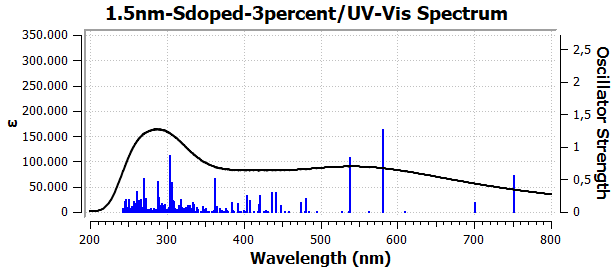} 
\end{figure*}

\begin{figure*}
\hspace*{-1cm}
\includegraphics[width=18cm,keepaspectratio]{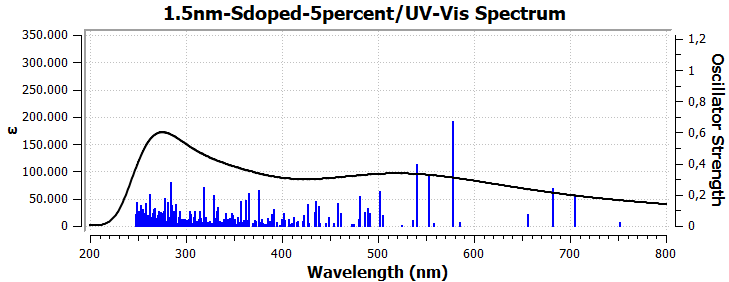} 
\end{figure*}

\begin{figure*}
\hspace*{-1cm}
\includegraphics[width=18cm,keepaspectratio]{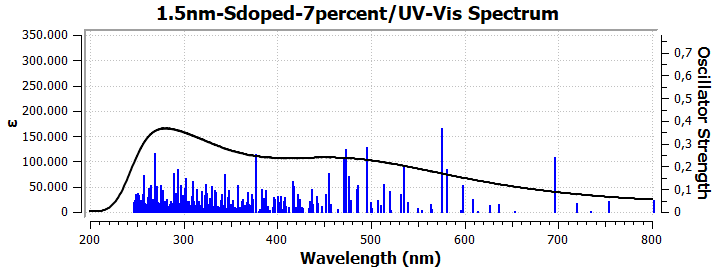} 
\end{figure*}

\begin{figure*}
\hspace*{-1cm}
\includegraphics[width=18cm,keepaspectratio]{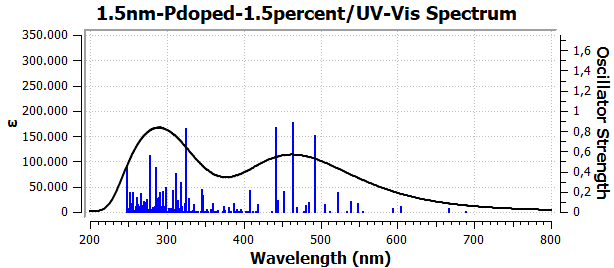} 
\end{figure*}

\begin{figure*}
\hspace*{-1cm}
\includegraphics[width=18cm,keepaspectratio]{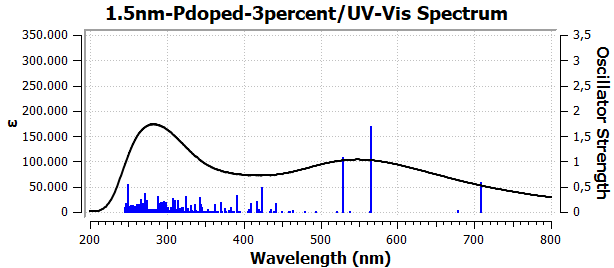} 
\end{figure*}

\begin{figure*}
\hspace*{-1cm}
\includegraphics[width=18cm,keepaspectratio]{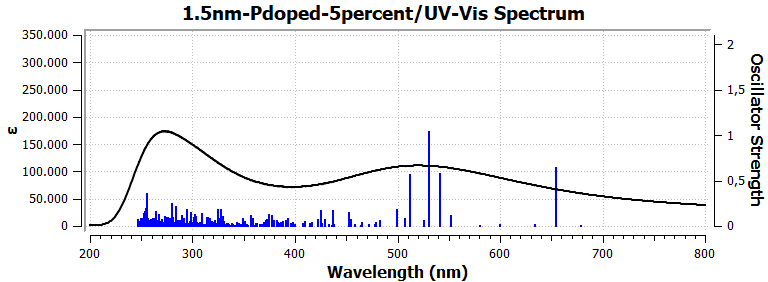} 
\end{figure*}

\begin{figure*}
\hspace*{-1cm}
\includegraphics[width=18cm,keepaspectratio]{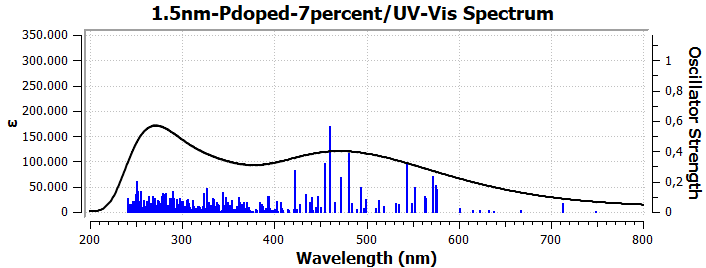} 
\end{figure*}

%2nm

\begin{figure*}
\hspace*{-1cm}
\includegraphics[width=18cm,keepaspectratio]{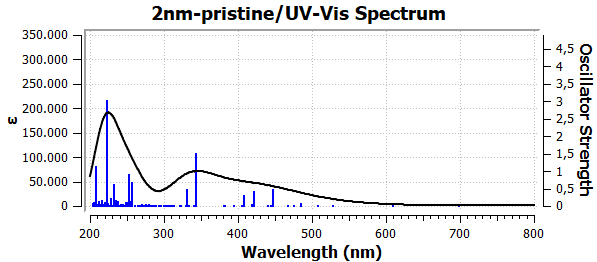} 
\end{figure*}

\begin{figure*}
\hspace*{-1cm}
\includegraphics[width=18cm,keepaspectratio]{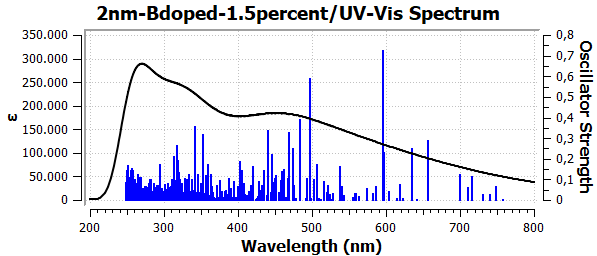} 
\end{figure*}

\begin{figure*}
\hspace*{-1cm}
\includegraphics[width=18cm,keepaspectratio]{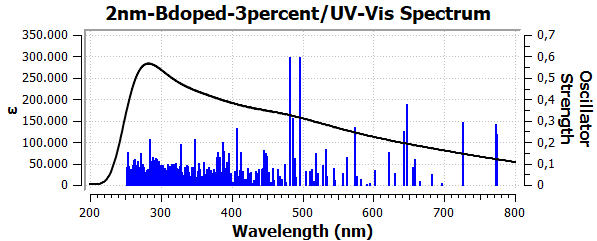} 
\end{figure*}

\begin{figure*}
\hspace*{-1cm}
\includegraphics[width=18cm,keepaspectratio]{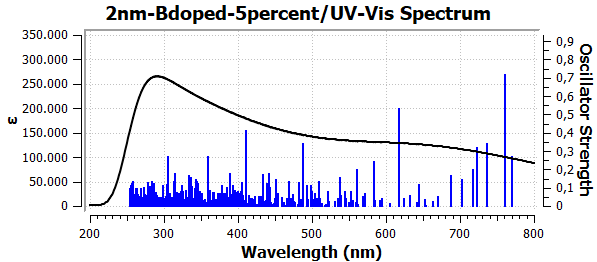} 
\end{figure*}

\begin{figure*}
\hspace*{-1cm}
\includegraphics[width=18cm,keepaspectratio]{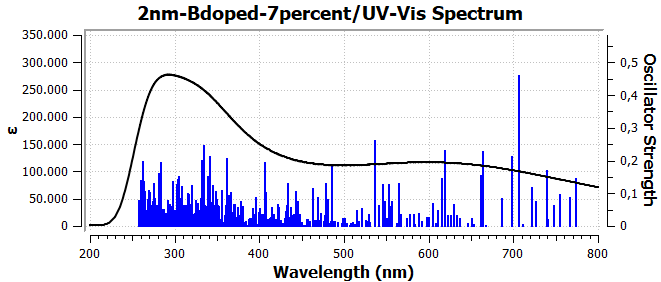} 
\end{figure*}

\begin{figure*}
\hspace*{-1cm}
\includegraphics[width=18cm,keepaspectratio]{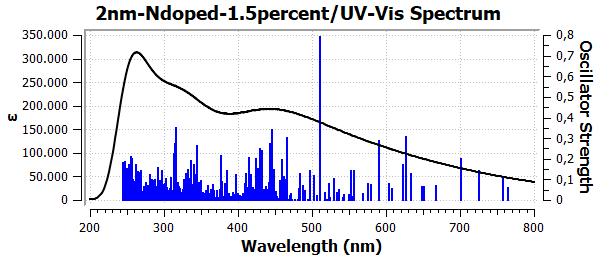} 
\end{figure*}

\begin{figure*}
\hspace*{-1cm}
\includegraphics[width=18cm,keepaspectratio]{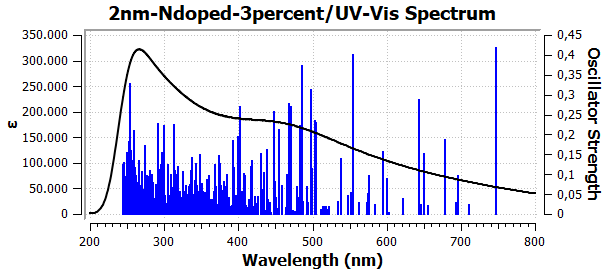} 
\end{figure*}

\begin{figure*}
\hspace*{-1cm}
\includegraphics[width=18cm,keepaspectratio]{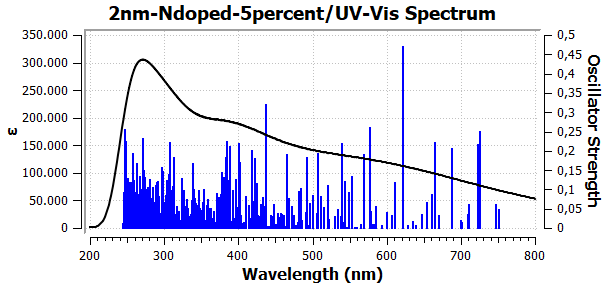} 
\end{figure*}

\begin{figure*}
\hspace*{-1cm}
\includegraphics[width=18cm,keepaspectratio]{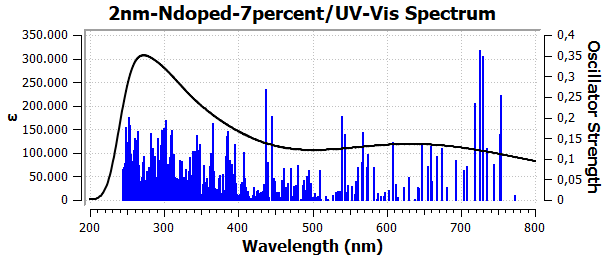} 
\end{figure*}

%https://tex.stackexchange.com/questions/46512/too-many-unprocessed-floats
\clearpage

\begin{figure*}
\hspace*{-1cm}
\includegraphics[width=18cm,keepaspectratio]{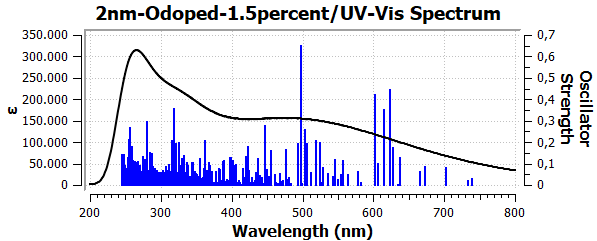} 
\end{figure*}

\begin{figure*}
\hspace*{-1cm}
\includegraphics[width=18cm,keepaspectratio]{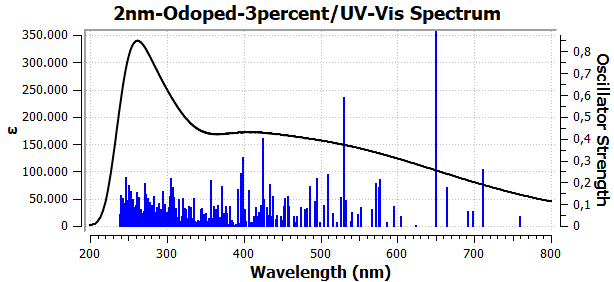} 
\end{figure*}

\begin{figure*}
\hspace*{-1cm}
\includegraphics[width=18cm,keepaspectratio]{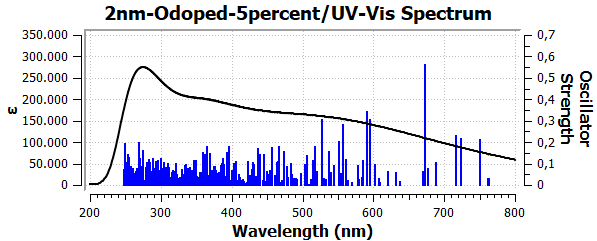} 
\end{figure*}

\begin{figure*}
\hspace*{-1cm}
\includegraphics[width=18cm,keepaspectratio]{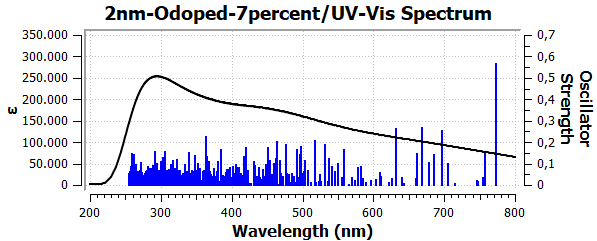} 
\end{figure*}

\begin{figure*}
\hspace*{-1cm}
\includegraphics[width=18cm,keepaspectratio]{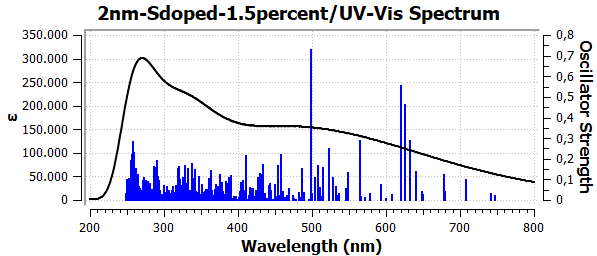} 
\end{figure*}

\begin{figure*}
\hspace*{-1cm}
\includegraphics[width=18cm,keepaspectratio]{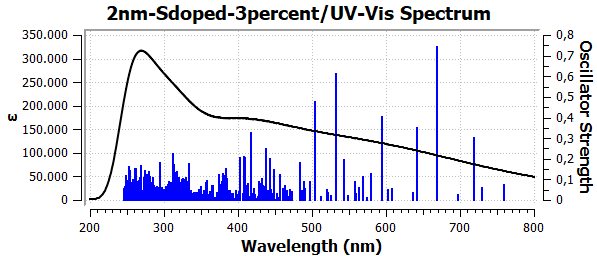} 
\end{figure*}

\begin{figure*}
\hspace*{-1cm}
\includegraphics[width=18cm,keepaspectratio]{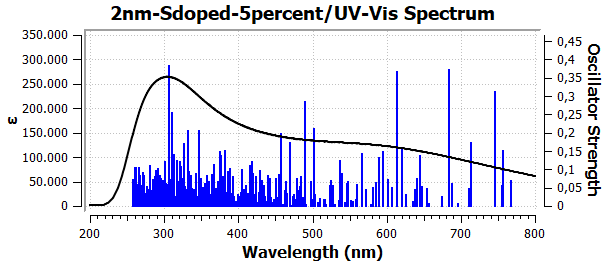} 
\end{figure*}

\begin{figure*}
\hspace*{-1cm}
\includegraphics[width=18cm,keepaspectratio]{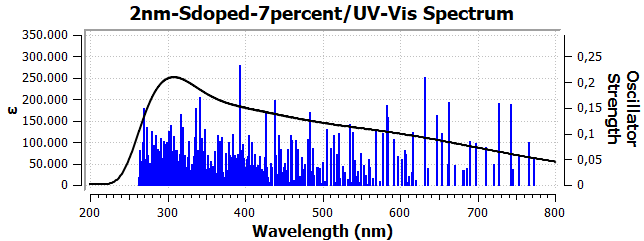} 
\end{figure*}

\begin{figure*}
\hspace*{-1cm}
\includegraphics[width=18cm,keepaspectratio]{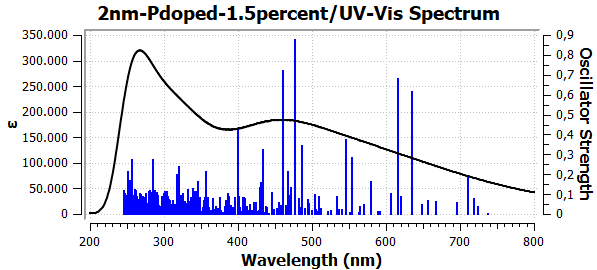} 
\end{figure*}

\begin{figure*}
\hspace*{-1cm}
\includegraphics[width=18cm,keepaspectratio]{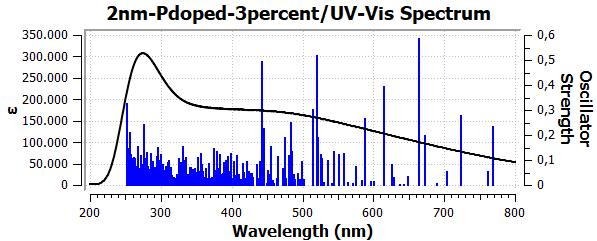} 
\end{figure*}

\begin{figure*}
\hspace*{-1cm}
\includegraphics[width=18cm,keepaspectratio]{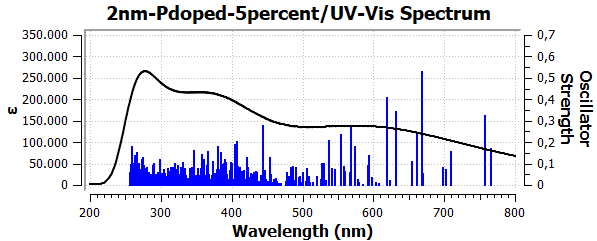} 
\end{figure*}

\begin{figure*}
\hspace*{-1cm}
\includegraphics[width=18cm,keepaspectratio]{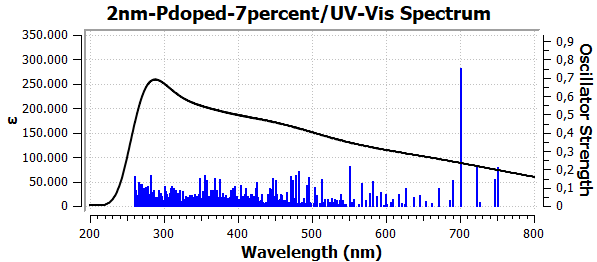} 
\end{figure*}

\end{document}